# The quark-gluon plasma, turbulence, and quantum mechanics

Mark Davidson
Spectel Research Corp., Palo Alto, CA 94303
Email: mdavid@spectelresearch.com

November 25, 2008

## Abstract

Quark-gluon plasmas formed in heavy ion collisions at high energies are well described by ideal classical fluid equations with nearly zero viscosity. It is believed that a similar fluid permeated the entire universe at about three microseconds after the big bang. The estimated Reynolds number for this quark-gluon plasma at 3 microseconds is approximately $10^{19}$. The possibility that quantum mechanics may be an emergent property of a turbulent proto-fluid is tentatively explored. A simple relativistic fluid equation which is consistent with general relativity and is based on a cosmic dust model is studied. A proper time transformation transforms it into an inviscid Burgers equation. This is analyzed numerically using a spectral method. Soliton-like solutions are demonstrated for this system, and these interact with the known ergodic behavior of the fluid to yield a stochastic and chaotic system which is time reversible. Various similarities to quantum mechanics are explored.

## Introduction

Extensive analysis of collision data at the Relativistic Heavy Ion Collider (RHIC) has confirmed that a quark-gluon plasma has been formed in Au-Au collisions at 40 TeV center of mass energy, and that this plasma behaves surprisingly like an ideal classical fluid with nearly zero viscosity [1-5]. Various hydrodynamic models have been proposed to describe it, many owing their genesis to Landau's seminal hydrodynamic model [6-8]. It is believed that a similar quark-gluon plasma permeated the entire universe about three microseconds after the big bang. Therefore, all that we observe today may have its origin in an essentially classical ideal fluid. Aside from the many interesting questions this fact poses for particle physicists and astrophysicists, it offers a new and unexpected possibility for the stochastic interpretation of quantum mechanics. While many physicists are trying to explain how a classical fluid can be formed with hindsight from the laws of the standard model, is it not also conceivable that the logical connection might flow in the opposite direction? Could not the laws of physics that we observe today, and especially the quantum laws, somehow have their origin in a violently turbulent classical fluid past which the RHIC experiments have now provided the first glimpse of? The idea that quantum mechanics is related to fluid mechanics is due to Madelung [9]. This paper takes some exploratory steps to examine this idea.

The lack of an underlying reality to quantum mechanics remains a weakness of modern physics in the opinion of a sizable group of physicists, mathematicians, and philosphers. A derivation of quantum mechanics from relativistic inviscid turbulent fuid dynamics would go a long way to remedying this situation, especially now that the RHIC experiment shows that such a dynamical system appears to be the origin of all that exists.

Most physicists have no problem accepting that a classical correspondence principle ensures that many quantum systems behave classically in the appropriate macroscopic limit. What if there were a dual "Quantum Correspondence Principle", so that in some limit quantum theory emerged from a particular and very special classical one. This need not work for all classical theories, obviously it doesn't, but only for one which would be a candidate for describing our universe. Finding such a theory has so far proven to be extremely difficult. But perhaps that is a strength rather than a weakness. Today, there are too many possible physical theories which cannot be ruled out. Maybe theoretical physics has set the bar too low for itself by shrinking away from this ontological completeness issue of quantum mechanics, allowing an overabundance of candidate theories to select from.

In recent years, owing to mathematical breakthroughs in string theory, the landscape of theories has become enormous [10], and the only principle that seems available to distinguish the standard theory of particle physics from a myriad of alternatives is the anthropic principle. Perhaps the task of constructing relativistic quantum theories is simply too easy with modern mathematical techniques, and is not a sufficient compass for future developments. Perhaps physics should set itself the additional task of deriving quantum theory from a classical statistical theory like fluid mechanics. This is a much harder task, and should limit the landscape significantly. In fact, many believe that with "reasonableness" assumptions the task of deriving quantum mechanics becomes impossible [11,12]. However, the verdict is not unanimous [13,14].

The first question is where to begin as literature on turbulence is vast. In reviewing it extensively, this author was drawn to recent research on the inviscid Burgers equation, and in particular to some very interesting numerical results of simulations to this equation using Galerkin truncation [15-17]. This paper is mainly about extending these results and asking if they might be indications of an emerging quantum behavior. Other authors have noted a mathematical similarity between forced inviscid Burgers equation and quantum field theory [18,19]. Similarities between vortex turbulence theory and quantum mechanics have been noted previously as well [20], based on an incompressible fluid model related to the vortex sponge model of the 19$^{th}$ century [21]. The Burgers equation has also been used in a spontaneous wave function collapse model which is quite relevant to this paper [22].

We start from an ideal relativisitic fluid, and make a connection to the inviscid Burgers equation to show its relevance to the problem at hand. All of the numerical simulations to follow were performed using the differential equation solvers contained in the commercial product Matlab version 6.0. For all the results presented here the method of solution of the fluid differential equations didn't make any significant difference, although run-times were affected.

## Cosmic relativistic Reynolds number for the quark gluon plasma at 3 microseconds

It is non-trivial to generalize the concept of Reynolds number from the non-relativistic Navier Stokes equation to the case of relativistic dissipative hydrodynamics. An approximate estimate was made for the quark gluon plasma at RHIC by Romatschke [23] who argued that the dimensionless Reynolds number can be roughly approximated by

$$\text{Re}_{RHIC} \sim \frac{TL_{RHIC}}{(\eta/s)} \sim 48\pi \quad (1)$$

where the following values were used: $L_{RHIC} = 6 \; fm$ (the radius of a gold nucleus); T=200 MeV; $\eta$ is the shear viscosity parameter and s the entropy density with the measured value $\eta/s \sim \frac{1}{8\pi}$ in natural units. In cosmology, it is believed that at 3 microseconds the entire universe was filled with a quark-gluon plasma of about the same energy density as at RHIC. Consider the Reynolds number for the universe then. The only parameter that would certainly be significantly different from RHIC would be the length scale. What should we take as the length scale at 3 microseconds for the universe? A logical first guess might be the distance to the causal horizon, ie. $L_U = c \cdot 3\mu S \sim 900 \; m$, where a flat spacetime has been assumed. Then the estimated Reynolds number of the universe is

$$\text{Re}_U = \frac{L_U}{L_{RHIC}} \cdot 48\pi \sim 2.26 \times 10^{19} \quad (2)$$

This is a large value as Reynolds numbers go. The reader may question the validity of using the horizon value for $L_U$, but consider that the cosmic background radiation shows that the horizon has had time to achieve thermal equilibrium in all directions as is generally explained today by inflation theory. The same mechanism might be at work in the quark-gluon fluid of the early universe giving it an even larger effective length scale than the horizon. The main point here is that the Reynolds number of the early universe is probably and plausibly extremely large, and turbulence is to be expected. If our estimate of the Reynolds number is off by a few orders of magnitude, it doesn't change this essential conclusion.

In this paper we will consider ideal fluids only, and will ignore any thermodynamic internal degrees of freedom. We note that turbulent inviscid fluids have exhibited emergent viscous behavior in numerical simulation [24], and so the nonzero value for $\eta/s$ as measured at RHIC, already the smallest viscosity ever measured for any material, does not necessarily rule out the possibility of an even lower viscosity or even zero viscosity in some protofluid whose highly developed turbulent state is being observed at RHIC. We are interested here in exploring the possibility that turbulence in such a protofluid might lead to quantum behavior as an emergent and purely classical phenomenon. We choose to work with an ideal fluid for the simple reason that quantum mechanics is time reversible, and viscosity would lead inevitably to irreversibility in the statistical dynamics of any emergent theory. Moreover, quantum mechanics has persisted essentially unchanged for at least the time since the universe became transparent to light, or about 13 or 14 billions years. Therefore, our best chance for getting quantum behavior from a fluid is to set viscosity to zero.

## Relativistic ideal fluid mechanics and the inviscid Burgers equation

The relativistic equation for an ideal fluid in curved spacetime is

$$\nabla_\beta T^{\alpha\beta} = 0 \qquad (3)$$

where $\nabla_\beta$ is the covariant derivative and where the energy momentum tensor is given by

$$T^{\alpha\beta} = p g^{\alpha\beta} + (p+\rho) U^\alpha U^\beta, \; g^{00} = +1, U^\alpha U_\alpha = +1, x = (x^0, \mathbf{x}) \qquad (4)$$

$U$ is the proper 4-velocity of the fluid, $c=1$, $p$ the pressure, and $\rho$ the mass density. We have in mind an elementary fluid here with no internal degrees of freedom.

We will assume that the fluid is timelike. This is certainly true for all standard material fluids, but considering that we are looking for a mathematical model for quantum mechanics, it is not inconceivable that we might consider a fluid which has spacelike velocities, perhaps as a second component, since the Bell non-locality experiments suggest that any hidden variable model of quantum mechanics, granting certain logical assumptions, must be non-local. There's another reason why we might want to consider a superluminal fluid. A mechanism for rapid thermalization at RHIC is not known, and poses a fundamental problem for the interpretation of experiments there. Moreover, the cosmological fluid also seems to have thermalized too rapidly too, as evidenced by the horizon problem in cosmology. Could these phenomena be related? If they are, then inflation is not likely to be the explanation, since no inflation has been observed at RHIC (thankfully!). We will not consider superluminal fluids or fluid components here, but we acknowledge that this might be a topic for future consideration.

This equation is completed by an independent equation of mass conservation

$$\nabla_\mu \rho U^\mu = 0 \tag{5}$$

and often by a barotropic equation of state relating pressure and density. The hydrodynamic equations that are used to simulate the RHIC experiments are much more complex than these [25,26]. Normally, one would need to specify an equation of state relating density, pressure and other parameters, and would include viscous effects, gauge field forces, and kinetic transport effects. We are looking for a tractable and simple model to start looking for the emergence of quantum-like behavior here. Surveying the literature, one finds an enormous amount of work has been done on the inviscid Burgers equation. Some of this work seems very relevant to our task at hand. The Burgers theory ignores pressure and most of the complications of real fluids, and we shall do the same. Ignoring pressure then, the energy momentum tensor and equation of motion become

$$T^{\alpha\beta} = \rho U^\alpha U^\beta, \quad and \quad U^\alpha \nabla_\alpha U^\beta = 0 \tag{6}$$

where we have used the product law for covariant derivatives. The equation involves only the velocity field.

In cosmology, it is standard to impose the cosmological principle – that on large scales the universe is both homogeneous and isotropic. This leads to the Robertson-Walker metric of the form [27]

$$d\tau^2 = dt_c^2 - R^2(t_c)\left\{\frac{dr^2}{1-kr^2} + r^2 d\theta^2 + r^2 \sin(\theta)^2 d\phi^2\right\} \tag{7}$$

Where $t_c$ is called cosmic time. Observation favors a flat space, ie. k=0, which we shall assume here. Cartesian coordinates are then more convenient, and we have

$$d\tau^2 = dt_c^2 - R^2(t_c)\{dx^2 + dy^2 + dz^2\} \tag{8}$$

The fluid equation can be written

$$U^\alpha \left(\partial_\alpha U^\nu + \Gamma^\nu_{\alpha\beta} U^\beta\right) = 0 \tag{9}$$

It can be shown that all of the Christoffel symbols are proportional to the time derivative of R

$$\Gamma^\nu_{\alpha\beta} \propto \frac{dR(t)}{dt} \tag{10}$$

The author is not aware of any measurements of this time derivative for the universe at only 3 microseconds of age, the time of the quark-gluon plasma. Therefore, in the interest of simplicity, we shall simply ignore curvature and set the Christoffel symbols to zero. The fluid equations become simply then what they would be for Minkowski space (a rescaling of the time variable is needed for complete agreement)

$$U^\alpha \partial_\alpha U^\nu = 0 \tag{11}$$

This equation has an infinite number of invariants as can be easily seen. Although $\rho$ drops out of the equation of motion, it does not follow that $\rho$ is simply a constant. In cosmology, this form for the energy momentum tensor is called the cosmic dust model [28]. It is compatible with a flat Robertson-Walker metric for space-time, provided a cosmological constant is included. Therefore, the assumption of a flat space here is theoretically compatible with general relativity provided a cosmological constant term is allowed. This type of universe is generally associated with cosmic inflation and is called the lambda-CDM model [29].

Let the velocity field $U^\mu$ be a solution to (11) and consider any scalar function $f$ which is a solution to the conservation equation (it need not correspond to a physical quantity)

$$\partial_\mu f U^\mu = 0 \tag{12}$$

Using this conserved function we define an energy-momentum-like invariant

$$T_f^{\mu\nu} = f U^\mu U^\nu \tag{13}$$

which satisfies

$$\partial_\mu T_f^{\mu\nu} = 0 \tag{14}$$

and consequently the following integrals are all invariants

$$P_f^\mu = \int fU^0 U^\mu d^3x = \int f\sqrt{1+\mathbf{U}^2}\, U^\mu d^3x \tag{15}$$

The fluid vorticity tensor is defined by

$$\omega_{\mu\nu} = \partial_\mu U_\nu - \partial_\nu U_\mu \tag{16}$$

Using the fluid equation (11) we find that

$$U^\mu \omega_{\mu\nu} = 0 \tag{17}$$

As a consequence of this equation the determinant of $\omega$ treated as a matrix must vanish. The situation here is similar to the force-free electromagnetic plasma theory of Uchida [30] where the vorticity tensor here plays the role of the electromagnetic tensor F. The determinant of $\omega$ is calculated by the Pfaffian

$$\det(\omega) = {}^*\omega^{\mu\nu}\omega_{\mu\nu}, \quad {}^*\omega^{\mu\nu} = \varepsilon^{\mu\nu\alpha\beta}\omega_{\alpha\beta} \tag{18}$$

where $\varepsilon$ is the totally antisymmetric tensor. One finds also

$$\partial_\lambda \omega_{\mu\nu} + \partial_\mu \omega_{\nu\lambda} + \partial_\nu \omega_{\lambda\mu} = 0 \tag{19}$$

$$\partial^\mu {}^*\omega_{\mu\nu} = 0 \tag{20}$$

Since the rank of a skew symmetric matrix must be even, it follows that the rank of $\omega$ must be 2 if it is nonzero. Thus $\omega$ has a two dimensional space of zero eigenvectors by the rank-nullity theorem. It also follows by analogy with [30] that $\omega^{\mu\nu}\omega_{\mu\nu} > 0$ and

$$\omega_{\mu\nu} = \partial_\mu \phi_1 \partial_\nu \phi_2 - \partial_\nu \phi_1 \partial_\mu \phi_2 \tag{21}$$

where $\phi_1$ and $\phi_2$ are scalar functions called Euler potentials.

The characteristic curves for this theory are simply straight trajectories for this particular model

$$\frac{dx^\mu(\tau)}{d\tau} = U^\mu(x^\mu(\tau)), \quad \frac{d^2x^\mu(\tau)}{d\tau^2} = 0 \rightarrow x^\mu(\tau) = x^\mu(0) + \tau v^\mu \tag{22}$$

Let us define a proper time for each spacetime point of the fluid in the following way. We have in mind here more general fluids than the simple cosmic dust case. The following assumes that the fluid's velocity is timelike and smooth. First of all, at $x^0 = 0$ in some Lorentz frame which we call the laboratory frame, we arbitrarily set the proper time $\tau$ for each point equal to zero. To calculate the proper time at other times in this frame, we use the characteristic curves defined by

$$\frac{d}{d\tau}x^\mu(\tau) = U^\mu(x(\tau)) \rightarrow x^\mu(\tau) = x^\mu(\tau_1) + \int_{\tau_1}^{\tau} U^\mu(x(\tau'))d\tau' \quad (23)$$

which can be re-expressed in terms of the time variable $x^0$

$$\frac{d}{dx^0}\mathbf{x}(x^0) = \frac{\mathbf{U}(\mathbf{x}(x^0), x^0)}{U^0(\mathbf{x}(x^0), x^0)} = \frac{\mathbf{U}(\mathbf{x}(x^0), x^0)}{\sqrt{1+\mathbf{U}(\mathbf{x}(x^0), x^0)^2}} \quad (24)$$

From the family of solutions to this, we can calculate the proper time at any point $(x^0, \mathbf{x})$ using

$$\tau(\mathbf{x}(x^0), x^0) = -\int_{x^0}^{0} \sqrt{1-(d\mathbf{x}/dt')^2}\, dt', \ x^0 \geq 0 \quad (25)$$

This assumes of course that the integrals exist. The only property we have used is the timelike property of the fluid velocity in order to define this proper time. The surfaces of constant $\tau$ are spacelike, but not planar in general. And so we now can consider the fluid as a function of $(\mathbf{x}, x^0)$ or of $(\mathbf{x}, \tau)$. This proper time should not be confused with the cosmic time in the Robertson-Walker metric since the velocity field of the fluid is not due solely to the expansion of the universe, but can be wildly complex and turbulent.

The reader might think at this stage that this theory is too trivial to be interesting. This is not the case because when one does a Fourier-Galerkin truncation the theory becomes non-integrable and incredibly rich [15-17]. The next step is to express the fluid's equation in terms of $\mathbf{x}$ and $\tau$ by using the partial derivative relation

$$\left.\frac{\partial}{\partial x^0}\right|_x = \frac{1}{U^0}\left.\frac{\partial}{\partial \tau}\right|_x \quad (26)$$

and so the fluid equation becomes

$$\frac{\partial}{\partial \tau}\mathbf{U}(\mathbf{x}, \tau) + (\mathbf{U}(\mathbf{x}, \tau) \cdot \nabla)\mathbf{U}(\mathbf{x}, \tau) = 0, \ U^0 = \sqrt{1+\mathbf{U}^2} \quad (27)$$

This is the three dimensional inviscid Burgers equation. It looks like a non-relativistic equation as it is Galilean covariant instead of Lorentz covariant, but as we've shown, it is actually a relativistic equation in disguise. The proper velocities can have magnitudes up to infinity, just as if it were a Galilean system. Note well though that an equal $\tau$ surface is a spacelike hypersurface, and not generally equal to an equal time surface except in very special circumstances. In turbulent fluid motion, this spacelike hypersurface will even be stochastic. This fact does not prevent us from treating the Burgers form of the equation like an ordinary Galilean invariant partial differential equation. So we will next consider some numerical results of this equation.

We will use simply t for proper time for the remainder, but ask the reader to remember that this corresponds to proper time as defined above, so that the equations may be interpreted as relativistic ones if desired, at least so long as shocks don't cause the transformation to be ill-defined.

**Background on the Burgers Equation**

The Burgers equation was first proposed as a model for zero pressure gas dynamics [31]. A review of the subject with applications is given in [32]. It is known to be integrable and therefore not a candidate for turbulence.

The Galerkin truncated version of this equation has been shown to have a number of interesting properties in numerical experiments [15-17,33] including ergodic chaotic behavior mimicking turbulence (see table 1). Because there is no viscosity in the inviscid theory, this turbulent behavior is qualitatively different from viscous turbulence in real fluids where continuous mixing is required to sustain the turbulence against the energy loss due to viscosity, and the viscosity leads to an energy cascade and equilibrium behavior often well approximated by Kolmogorov scaling [34] in a range of wave numbers called the inertial range. Three dimensional simulations of truncated, incompressible, and inviscid Euler equations have shown similar chaotic turbulent results [24], including an interesting transient Kolmogorov scaling regime resembling a viscous fluid caused by the flow of energy from large scales to small as required by ergodic behavior. This result shows that some aspects of viscosity can emerge from an inviscid fluid theory in a chaotic turbulent regime.

| |
|---|
| The fluid equations are time reversible, Galilean invariant, and parity invariant |
| The fluid equation is invariant under a constant scaling of time |
| There are three conserved quantities, $u_0$, $E$, and $H$ |
| The equilibrium distribution is invariant under time reversal and parity if $H=0$ |
| The equilibrium distribution violates time reversal and parity if $H \neq 0$ |
| The solutions are ergodic for most starting conditions |
| There are invariant subspaces |
| The energy spectrum in equilibrium exhibits equipartition when $H=0$ |
| The energy spectrum in equilibrium is tilted when $H \neq 0$ |

**Table 1** Properties of the 1D Fourier-Galerkin truncated inviscid Burgers Equation

The truncated inviscid Burgers equation is an extremely simple one-dimensional nonlinear model which nevertheless shares extraordinarily complex features with more realistic but much more complex continuum systems. In [16] it was shown that there are three conserved quantities for this system, and that for many randomly selected starting conditions the equations are chaotic and ergodic and result in equipartition of energy, but also that for certain non-typical starting conditions, the system does not result in equipartition, but rather has a tilted energy spectrum. This correlates well with the relative magnitudes of two of the conserved quantities, the traditional energy being one, and the other being a third order sum which is referred to in the literature as the Hamiltonian.

We shall show in this paper that there are soliton solutions to this system of equations, and that they are related to the deviation from equipartition, and to extremal values of the Hamiltonian. These soliton solutions are similar to the delta solitons proposed and analyzed by Sarrico [35] which were inspired by the seminal works of Maslov et al. [36,37].

**The Inviscid Burgers Equation in one dimension**

The continuum version of the equation (also called the Burgers-Hopf equation) is

$$\frac{\partial u}{\partial t} + \frac{1}{2}\frac{\partial}{\partial x}(u^2) = 0 \tag{28}$$

We use the symbol $t$ for time here to conform to standard notation on this subject, even though it is actually proper time in a relativistic theory. This can be interpreted as a pressureless Euler equation for an ideal fluid in one dimension where $u$ is the velocity. We shall impose periodic boundary conditions so that $u(x+2\pi,t) = u(x,t)$. Using the scaling properties of this equation, it is a simple matter to rescale the periodic distance to any number one desires.

A non-canonical Hamiltonian can be defined which generates the equations of motion [16]

$$H = \frac{1}{12\pi}\int_0^{2\pi} u^3 dx \tag{29}$$

The equations of motion are written

$$u_t = \left(-2\pi\frac{\partial}{\partial x}\right)\frac{\delta H}{\delta u} \tag{30}$$

The mean energy density as defined in [15-17,33] is a conserved quantity.

$$E = \frac{1}{4\pi}\int_0^{2\pi} u^2 dx, \quad \frac{dE}{dt} = 0 \tag{31}$$

The total energy is simply $2\pi$ times this. Also, H is a conserved quantity $H_t = 0$. In fact, any function of the form $\int_0^{2\pi} g(u)dx$ is also an invariant provided g is differentiable [16].

In order to understand the nature of the solitons that we will find in the truncated case, we consider a variational problem. Let H be given, and let us choose $u$ so that the energy $E$ is extremal subject to the constraint imposed by H fixed. We use a Lagrange multiplier technique

$$S = H + \lambda E, \quad \frac{\delta S}{\delta u} = 0, \quad \text{where } \lambda \text{ is a Lagrange multiplier} \tag{32}$$

$$\frac{\delta H + \lambda E}{\delta u} = 0 = \frac{1}{4\pi}u^2 + \frac{1}{2\pi}\lambda u \tag{33}$$

and so either $u = 0$ or $u = -2\lambda$. The solutions are either bivalued and discontinuous or simply constant in x. We shall see that for the truncated theory, this same variational principle has nontrivial solutions including solitons. The inviscid Burgers equation is invariant under a simultaneous Galilean transformation, space translation, and time translation $u(x,t) \Rightarrow u(x+a+\mathrm{v}t, t+\tau) - \mathrm{v}$, where $a, \mathrm{v}$, and $\tau$ are constants. It is invariant under time scaling $u(x,t) \Rightarrow \alpha u(x, \alpha t)$. It is also invariant under parity and time reversal.

**Fourier-Galerkin truncation**

$P_\Lambda f = f_\Lambda$ denotes the Fourier projection operator with cutoff $\Lambda$

$$P_\Lambda f = f_\Lambda = \sum_{|k|<=\Lambda} \hat{f}_k e^{ikx} \tag{34}$$

$$\hat{f}_k = \frac{1}{2\pi} \int_0^{2\pi} f(x) e^{-ikx} dx \tag{35}$$

where k is integer, $f(x)$ is $2\pi$ periodic and real-valued so that $\hat{f}_{-k} = \left(\hat{f}_k\right)^*$. The truncated Burgers-Hopf or inviscid Burgers equation is [15-17,33]

$$\frac{\partial u_\Lambda}{\partial t} + \frac{1}{2}\frac{\partial}{\partial x} P_\Lambda(u_\Lambda^2) = 0 \tag{36}$$

the energy density becomes

$$E = \frac{1}{4\pi} \int_0^{2\pi} P_\Lambda(u_\Lambda^2) dx \tag{37}$$

$$E = \frac{1}{2}\sum_{|k|<=\Lambda} \hat{u}_k \hat{u}_{-k} = \frac{1}{2}\sum_{|k|<=\Lambda} |\hat{u}_k|^2 = \frac{|\hat{u}_0|^2}{2} + \sum_{k=1}^{\Lambda} |\hat{u}_k|^2 \tag{38}$$

and the Hamiltonian becomes

$$H = \frac{1}{12\pi} \int_0^{2\pi} P_\Lambda(u_\Lambda^3) dx \tag{39}$$

it can be shown that

$$\frac{d}{dt}u_0 = \frac{dE}{dt} = \frac{dH}{dt} = 0 \tag{40}$$

and that

$$H = \frac{1}{6} \sum_{\substack{k_1+k_2+k_3=0 \\ |k_1|,|k_2|,|k_3|<=\Lambda}} \hat{u}_{k_1} \hat{u}_{k_2} \hat{u}_{k_3} \tag{41}$$

$$\frac{\partial \hat{u}_k}{\partial t} = \frac{-ik}{2} \sum_{|k'|,|k-k'|<=\Lambda} \hat{u}_{k-k'} \hat{u}_{k'} = -ik\frac{\partial H}{\partial u_k^*} \tag{42}$$

this is left invariant by the Galilean transformation $\hat{u}_0 \to \hat{u}_0 - v, \hat{u}_k \to \hat{u}_k e^{ikvt}$, and by space and time translations, parity and time reversal.

**Extremal condition and solitons**

We now consider making E extremal subject to the constraint imposed by holding H fixed. Again we use the Lagrange multiplier method. To simplify, let's work in the Galilean frame of reference where $\hat{u}_0 = 0$. There is no loss of generality in doing this due to Galilean invariance.

$$\frac{\partial}{\partial \hat{u}_k}[H + \lambda E] = 0, k \neq 0 \tag{43}$$

$$\frac{\partial}{\partial \hat{u}_k}\left[\frac{1}{6}\sum_{\substack{k_1+k_2+k_3=0 \\ |k_1|,|k_2|,|k_3|<=\Lambda}} \hat{u}_{k_1}\hat{u}_{k_2}\hat{u}_{k_3} + \lambda\left(\frac{1}{2}\sum_{|k'|<=\Lambda}\hat{u}_{k'}\hat{u}_{-k'}\right)\right] = 0, k \neq 0 \tag{44}$$

We can do the variation so that the reality condition is preserved, which implies $\delta\hat{u}_k = \delta\hat{u}_{-k}^*$ or we can vary the $\hat{u}_k$ and $\hat{u}_{-k}$ independently which is a stronger condition and simpler to analyze. We choose the later and so obtain

$$\left[\frac{1}{2}\sum_{\substack{k+k_2+k_3=0 \\ |k_2|,|k_3|<=\Lambda}} \hat{u}_{k_2}\hat{u}_{k_3} + \lambda\hat{u}_{-k}\right] = 0, 0 < |k| <= \Lambda \tag{45}$$

which can be rewritten

$$\left[\frac{1}{2}\sum_{|k_2|,|k-k_2|<=\Lambda} \hat{u}_{k_2}\hat{u}_{k-k_2} + \lambda\hat{u}_k\right] = 0, 0 < |k| <= \Lambda \tag{46}$$

Now use (42) to obtain

$$\frac{\partial \hat{u}_k}{\partial t} - ik\lambda\hat{u}_k = 0 \tag{47}$$

which is equivalent to

$$\frac{\partial u(x,t)}{\partial t} - \lambda\frac{\partial u(x,t)}{\partial x} = 0 \tag{48}$$

and this has solutions of the form

$$u(x,t) = u(x + \lambda t) \tag{49}$$

And therefore the extremal solutions are traveling waves (with the direction determined by the sign of $\lambda$), provided that (46) has non-trivial solutions. From (46) we obtain a value for $\lambda$ after multiplying by $u_{-k}$ and summing over $k \neq 0$

$$\lambda = -\frac{3H}{2E} \tag{50}$$

This equation is valid only in the frame at which the mean fluid velocity is zero. The wave velocity is just $-\lambda$ in this frame.

Since the equations are invariant under Galilean transformations, we can slow down the traveling wave to zero velocity by moving along with it, and then the solution becomes just a static function of x. Note that $\hat{u}_0 \neq 0$ in this co-moving frame. This static solution must satisfy

$$\frac{\partial \hat{u}_k}{\partial t} = 0 = \frac{-ik}{2} \sum_{\substack{|k'|,|k-k'|\leq \Lambda \\ |k|\leq \Lambda}} \hat{u}_{k-k'} \hat{u}_{k'} \tag{51}$$

This equation is invariant under scale transformations, so that if $\hat{u}_k$ is a solution then so is $\sigma \hat{u}_k$ for constant $\sigma$ and all k. Separating out the $k'=0$ and $k'=k$ terms in the sum we obtain

$$\hat{u}_k \hat{u}_0 = -\frac{1}{2} \sum_{\substack{|k'|,|k-k'|\leq \Lambda \\ k',k-k'\neq 0}} \hat{u}_{k-k'} \hat{u}_{k'}, \quad \text{for all nonzero } k \text{ s.t. } 0<|k|\leq \Lambda \tag{52}$$

Now we can divide by $\hat{u}_0^2$ to obtain

$$\frac{\hat{u}_k}{\hat{u}_0} = -\frac{1}{2} \sum_{\substack{|k'|,|k-k'|\leq \Lambda \\ k',k-k'\neq 0}} \frac{\hat{u}_{k-k'}}{\hat{u}_0} \frac{\hat{u}_{k'}}{\hat{u}_0}, \quad \text{assuming that } \hat{u}_0 \neq 0 \text{ and } 0<|k|\leq \Lambda \tag{53}$$

This equation can be iterated, starting with a trial vector. Convergence is assisted if we solve instead

$$f_k = -\alpha(f) \frac{1}{2} \sum_{\substack{|k'|,|k-k'|\leq \Lambda \\ k',k-k'\neq 0}} f_{k-k'} f_{k'}, \quad 0<|k|\leq \Lambda \tag{54}$$

with the regularization factor $\alpha$ given by

$$\alpha(f) = \frac{1}{\text{maximum}\left(\left\{\left|-\frac{1}{2} \sum_{\substack{|k'|,|k-k'|\leq \Lambda \\ k',k-k'\neq 0}} f_{k-k'} f_{k'}\right|, k=-\Lambda \text{ to } \Lambda, k\neq 0\right\}\right)} \tag{55}$$

Then it follows that the desired solution is

$$\frac{\hat{u}_k}{\hat{u}_0} = \alpha f_k \tag{56}$$

Iteration proceeds as follows

$$f^{N+1}_k = -\alpha(f^N) \frac{1}{2} \sum_{\substack{|k'|,|k-k'|\leq \Lambda \\ k',k-k'\neq 0}} f^N_{k-k'} f^N_{k'}, \tag{57}$$

The starting point $f^0_k$ is arbitrary but nonzero. Different starting positions converge to different solutions. A proof has not been found that this algorithm always converges, but the practical experience with it is that it has converged to a soliton for almost all of the starting values studied by the author.

**Numerical examples showing soliton solutions**

In (53) let us set $\hat{u}_0 = 1$ without loss of generality since all other values can be obtained by a scale transformation. Figure 1 shows a static soliton obtained by setting all the $f^0_k = 1, 0<|k|\leq \Lambda$ and 0 otherwise in (57), solving the iteration numerically, and then translating the soliton peak to the position x=0. The iterative solution was terminated when the change in the absolute value of all the elements of $f$ were less than $10^{-12}$.

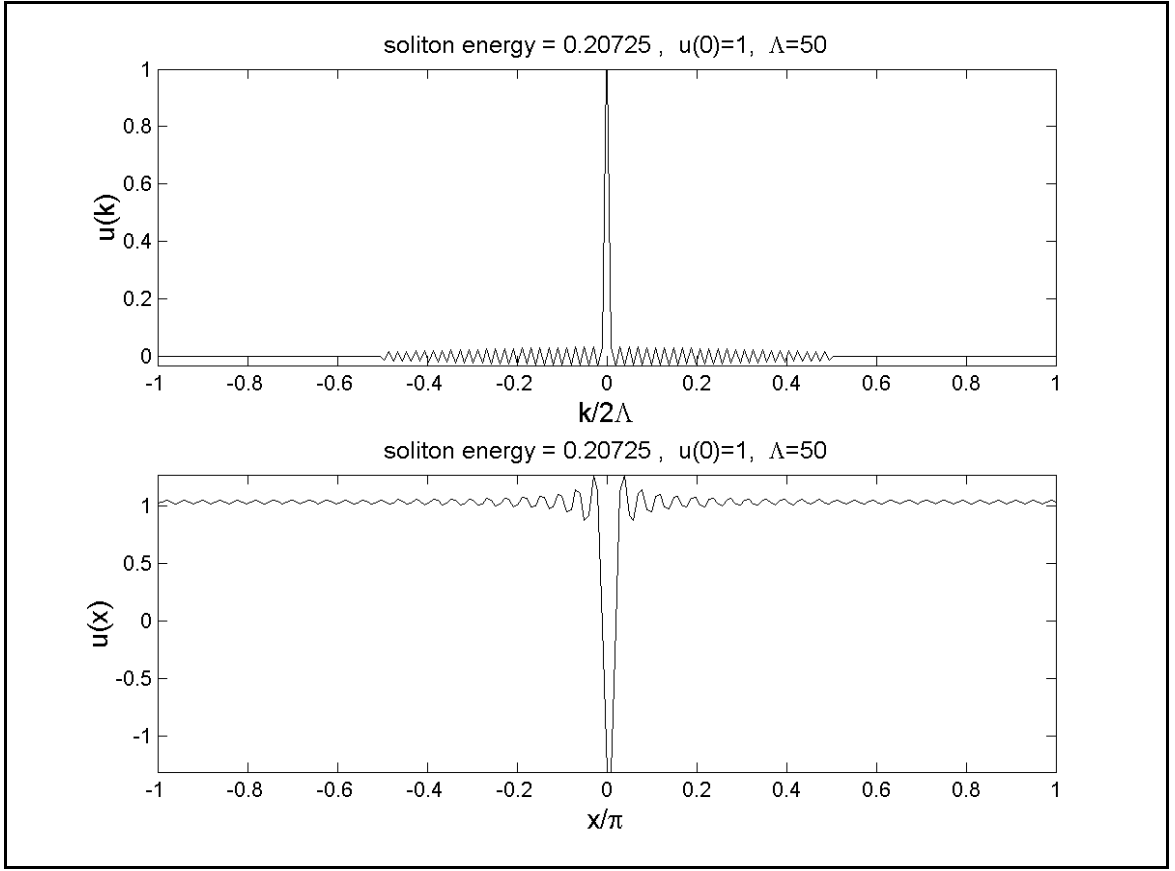

Fig. 1 A simple soliton. The energy reported in this figure is $2\pi E$ and in the rest frame of the soliton the fluid had a mean velocity of 1.

By multiplying the $\hat{u}_k$ by a constant scale factor, we can produce solitons of any velocity. The energy of the soliton is given by

$$E = 2\pi \sum_{k=1}^{\Lambda} |\hat{u}_k|^2 \qquad (58)$$

and therefore, in the rest frame of the background fluid ($\hat{u}_0 = 0$) it's energy will be proportional to its velocity squared which is just the result from Newtonian mechanics, where $E = \frac{1}{2}m\mathbf{v}^2$. This same Newtonian behavior was also found by Sarrico [35] for his delta solitons which appear to be closely related to the soliton solutions that we have found here. The following empirical formula fits the soliton data approximately.

$$\hat{u}_k = a + b\sin(2\pi k/(\Lambda d) + c), k = 1\ to\ \Lambda \qquad (59)$$

| Λ | a | b | c | d |
|---|---|---|---|---|
| 50 | -0.016819619 | 0.016090415 | 4.9485951 | 4.3598829 |
| 100 | -0.0083189075 | 0.0079554749 | 4.9485211 | 4.3374721 |
| 200 | -0.0041370769 | 0.0039556146 | 4.944829 | 4.326271 |
| 1000 | -0.00082386747 | 0.00078761299 | 4.948452 | 4.3173147 |
| 5000 | -0.00016463228 | 0.00015738281 | 4.9484457 | 4.3155235 |

**Table 2** Empirical fit parameters for soliton functions with different Fourier cutoff $\Lambda$

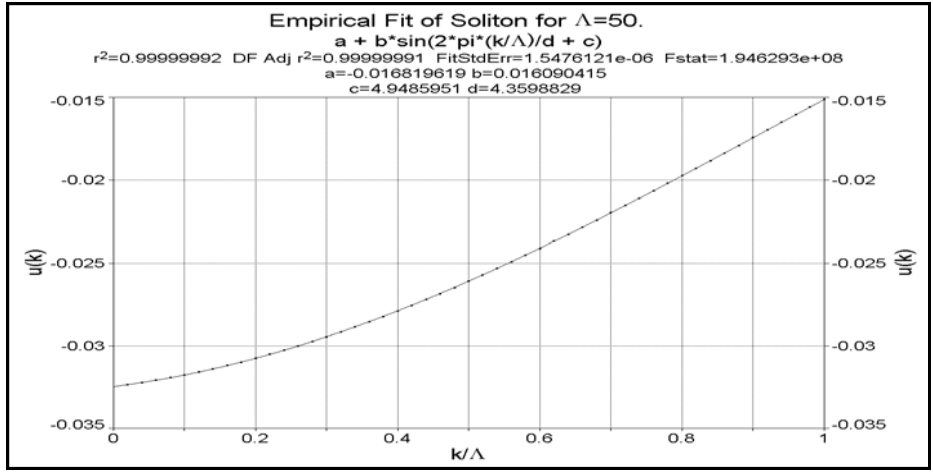

Fig. 2 An example of an empirical fit to the soliton function. The dots are the numerically calculated results, the curve the empirical fit

In table 2 is presented a numerical fit to the nonzero terms for the static soliton found for various orders $\Lambda$. From this data it is seen that the parameters a and b vary approximately as $1/\Lambda$ for large $\Lambda$, and that c and d are approximately constant. Therefore it follows that for large $\Lambda$ we have in the frame where the fluid's mean velocity is zero

$$E_{Soliton} = \frac{g}{\Lambda} + o(\frac{1}{\Lambda}), \text{ for some constant g} \tag{60}$$

Thus the energy of a single soliton with fixed velocity goes to zero as we take the cutoff to infinity, assuming of course that the trends shown in table 2 continue to larger cutoffs. This suggests that if we are to relate these solitons to actual particles, then since the cutoff would be very large, they would have to be quite low mass particles.

The soliton solutions have been tested with a differential equation solver which integrates the truncated equations using a Runga-Kutta ODE algorithm. These simulations confirm that the soliton solutions found are indeed behaving like solitons.

**Double soliton solutions**

The solutions to the static soliton equations shown in Figures 3 and 4 are double solitons. They have approximately twice the energy of the single soliton. These are found by trying different starting functions for the iteration, for example by starting the iteration with a superposition of two solitons displaced from one another. It is not clear how many solutions the equations have. The author has not found a three soliton solution yet. However, integrating these double soliton solutions in time shows that they do not behave exactly as solitons should. At first and for a period of time, the separation between the two solitons remains constant as it should, but after a time they move towards each other and eventually collide. The equations of the fluid are highly chaotic, and the error for turbulent fluid motion grows exponentially. This failure of the double soliton solutions to act as a soliton after an initial time has passed is probably due to the exponentially growing error of the solution. This suggests that the double solitons are unstable to small perturbations.

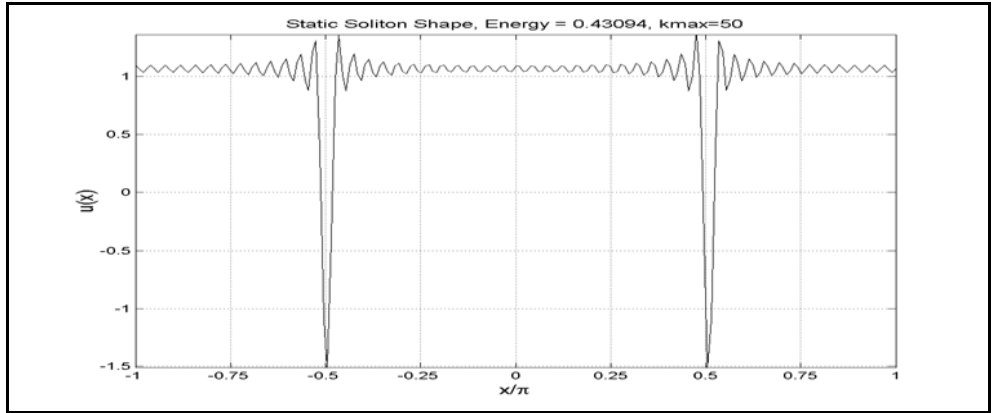

Fig. 3 A Double soliton. The energy reported in this figure is $2\pi E$.

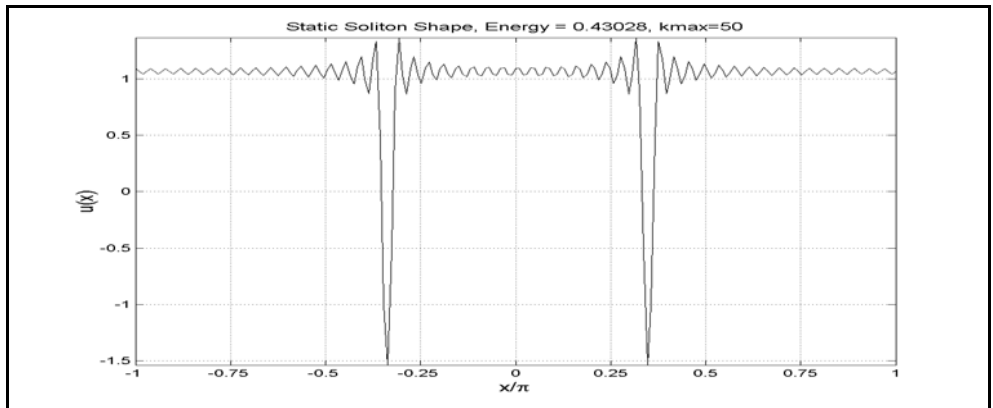

Fig. 4 A second double soliton. The energy reported in this figure is $2\pi E$

**Time series of a single soliton with noise**

In figure 5 we see a time series for a pure soliton. It moves without changing shape. In figure 6 we see the same soliton, but with additional random starting velocity noise field of the form

$$\hat{u}(k) = \frac{\sigma}{\sqrt{2}} X_k, \text{ k=1 to } \Lambda, \hat{u}(-k)=\hat{u}(k)^* \tag{61}$$

where $X_k$ is a normally distributed complex random variable whose real and imaginary parts have variance 1. This noise is equipartitioned on the average, and this mimics the equilibrium chaos of the typical starting conditions as found in [16]. The noise is simply added to the pure soliton solution for the initial conditions. The result shows that the soliton moves in a random and chaotic velocity field. Despite the chaotic background, the soliton maintains its identity and average velocity as time progresses without noticeable deterioration or slowing down seemingly forever, and it moves with approximately constant velocity slightly different from its unperturbed velocity. The velocity of the soliton is modified slightly by the presence of noise, even on the average. The peak position of the soliton shows a random diffusive motion in figure 7.

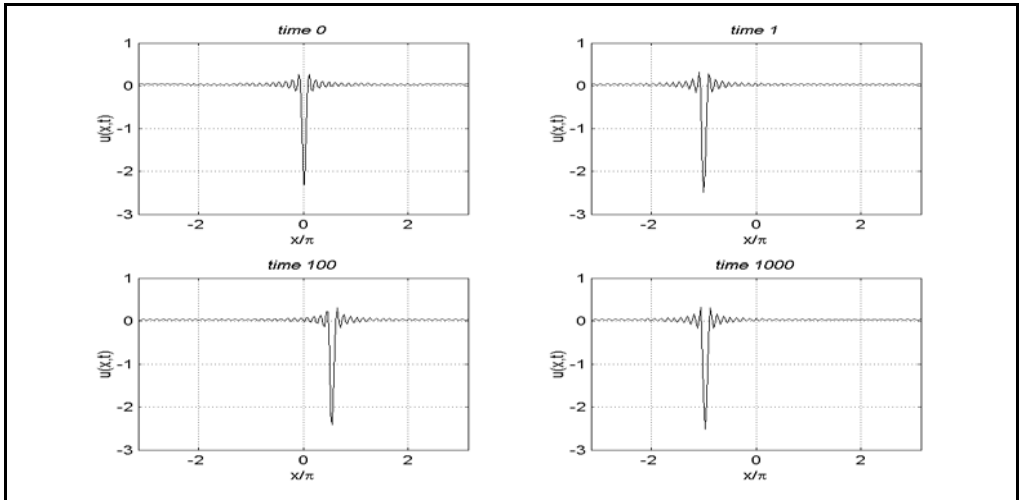

Fig. 5  A time series of a single moving soliton viewed in the frame where the mean fluid velocity is zero.  It maintains its shape as it moves to the left with soliton velocity of -1 radian/s

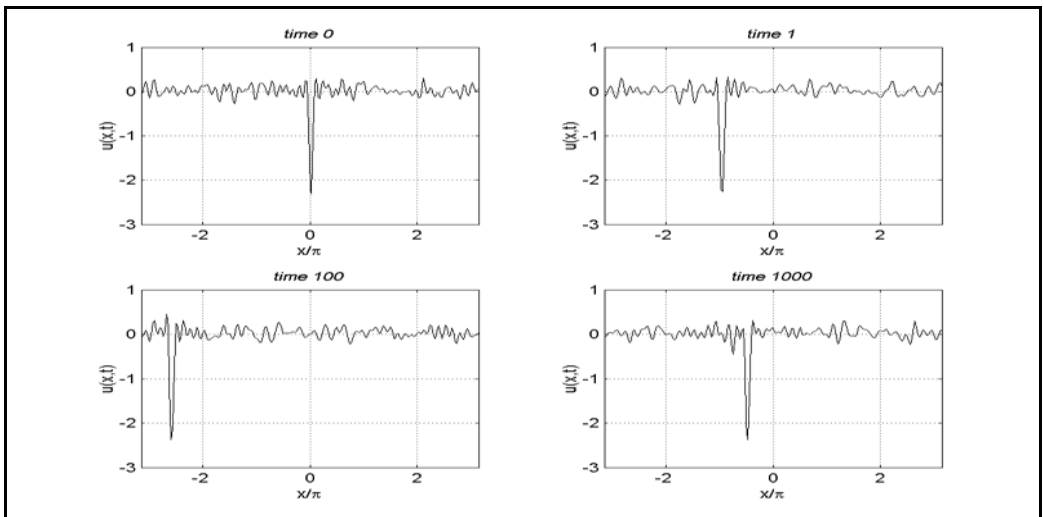

Fig. 6  A time series of a single moving soliton with random noise ($\sigma = .01$) initial starting condition as viewed in the frame where the mean velocity of the fluid is zero.  The soliton still maintains its identity over long times, and its motion has a random component due to its interaction with the background random turbulent fluid.  The background is observed to be approximately equipartitioned in k.

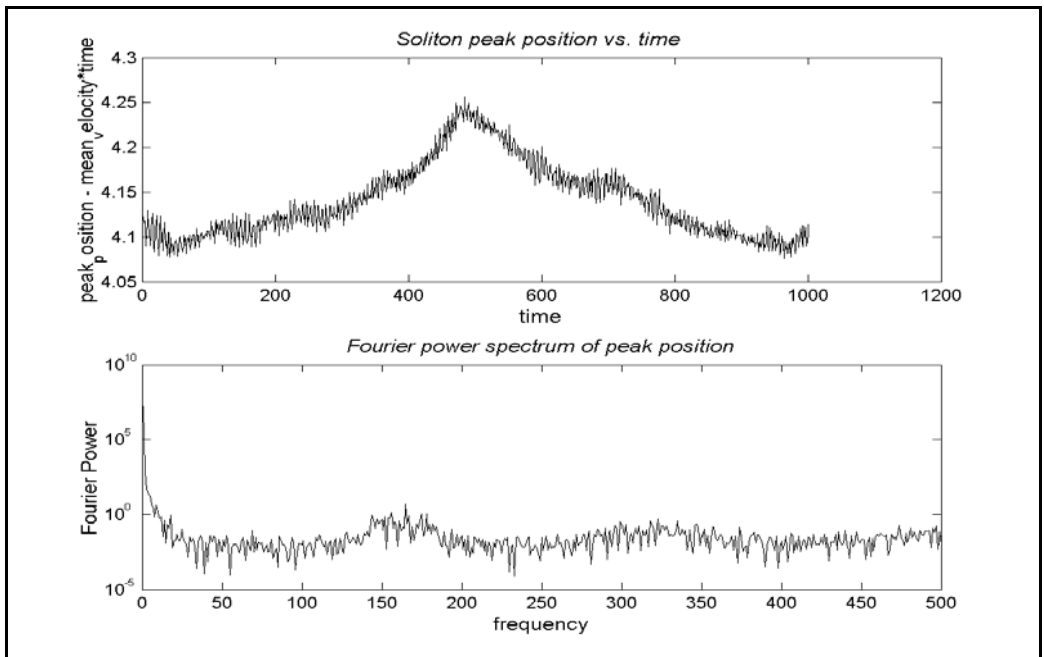

Fig. 7 The top graph shows the peak position of the soliton with noise after subtracting out the mean velocity motion, and the bottom graph shows the Fourier power spectrum of this trajectory.

These results are rather amazing. We have a system here which provides random diffusion of localized solitons without any viscosity. The system is time reversible and the diffusion persists forever. It's remarkably similar to quantum mechanics, yet the basic model is so simple.

**Colliding solitons**

Figures 8 and 9 show colliding solitons at different cutoff frequencies $\Lambda$. The solitons suffer considerable damage on collision, and this damage does not seem to be decreasing with increasing $\Lambda$. As a numerical check, the damaged soliton ending state was rerun with time reversed to see if the initial starting undamaged solitons reappeared, and they did. By some strict definitions of solitons, the fact that they are destroyed in collisions would mean they aren't truly solitons. But some looser definitions allow the term soliton to be applied with the term "indestructible soliton" reserved for the cases that pass through each other with only a phase shift. We call them simply solitons here in the interest of brevity.

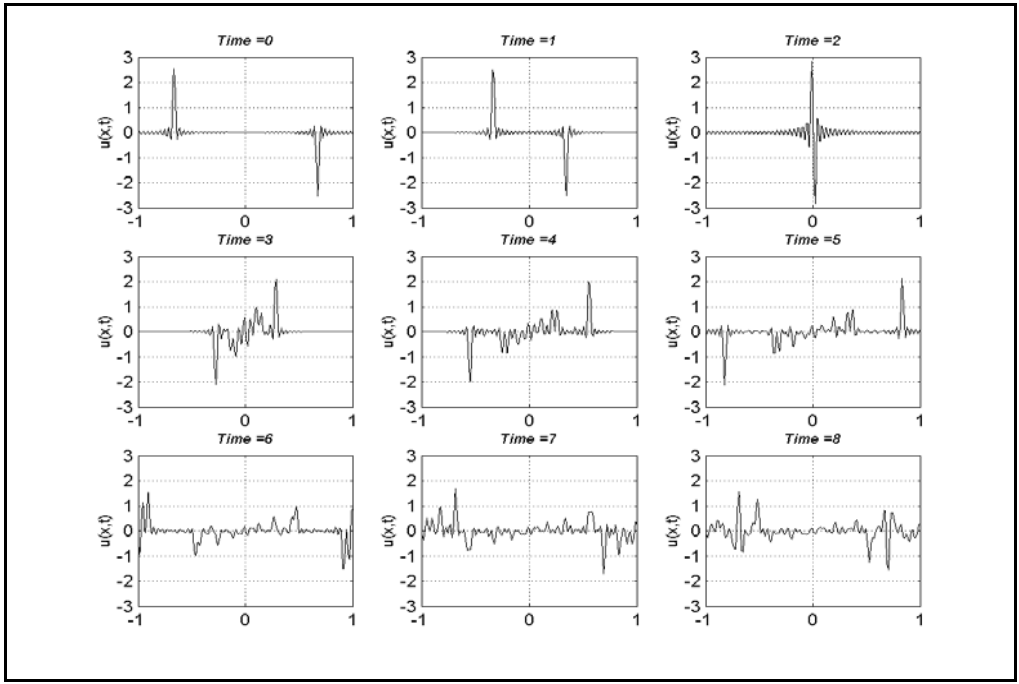

Fig. 8 Colliding solitons with equal and opposite velocity, $\Lambda = 50$

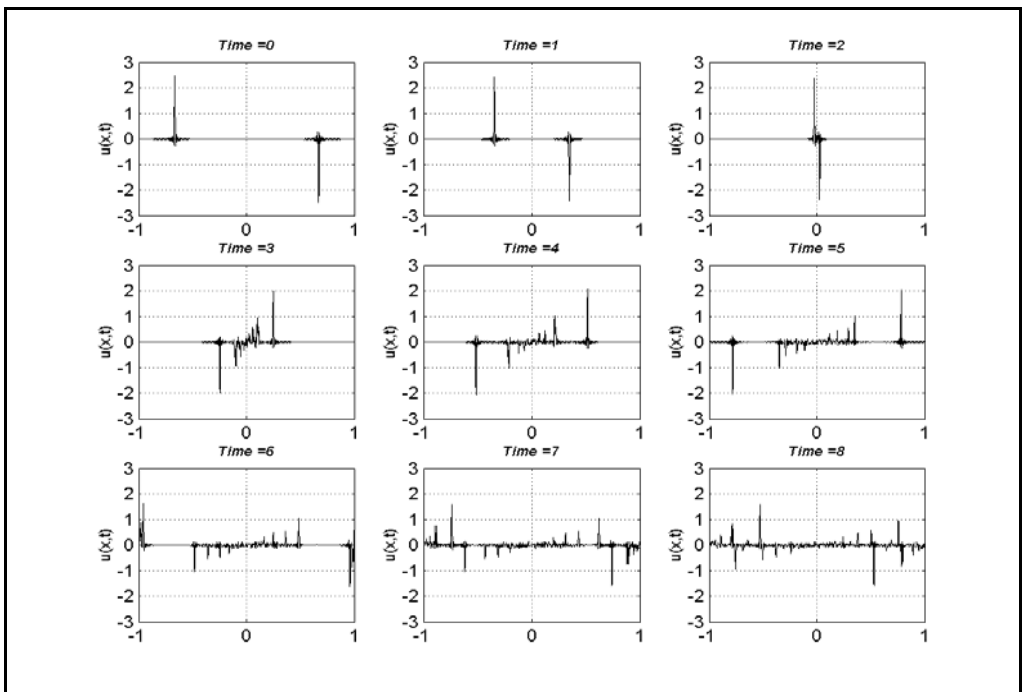

Fig. 9 Colliding solitons with equal and opposite velocity, $\Lambda = 200$

**Attractive force observed between two solitons moving with the same velocity**

We superimpose two identical solitons shifted by some amount d in x relative to one another. This is not at all the same problem as the double soliton solutions considered above. Here the two single soliton solutions are just superimposed. We definitely don't expect this superposition to act like a soliton. We add their velocity fields linearly, and integrate the fluid motion.

$$u(x,0) = u_s(x) + u_s(x-d) \tag{62}$$

A time series showing these two solitons moving towards each other, experiencing apparent attraction is shown in figure 10. The relative separation between the two peaks of these solitons is plotted in figure 11. After they collide, they experience damage.

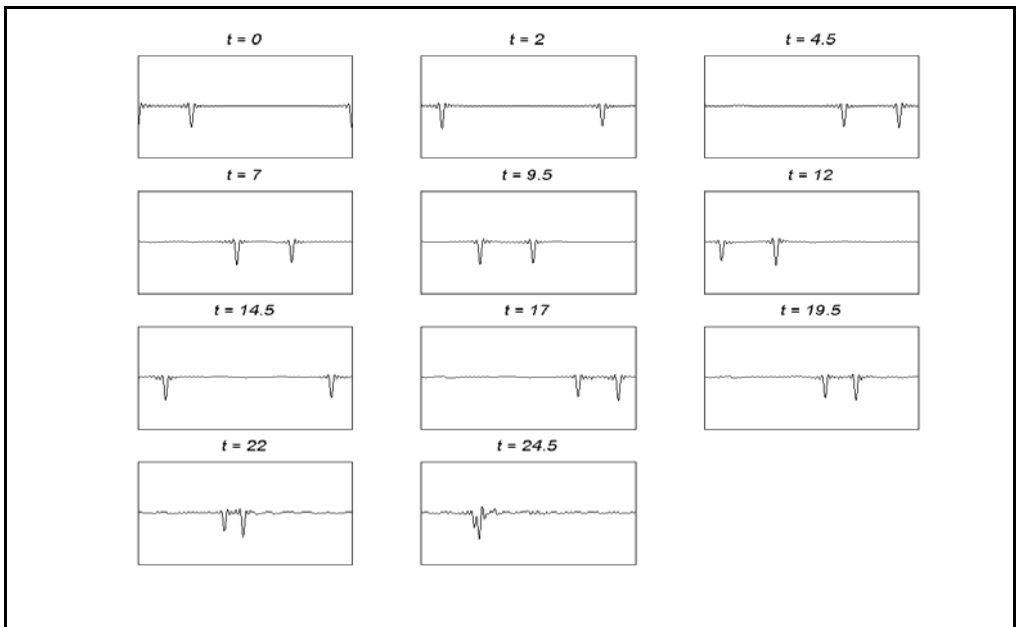

Fig. 10 A time series of two solitons moving to the left initially at the same velocity, but eventually experiencing an attraction and eventually colliding.

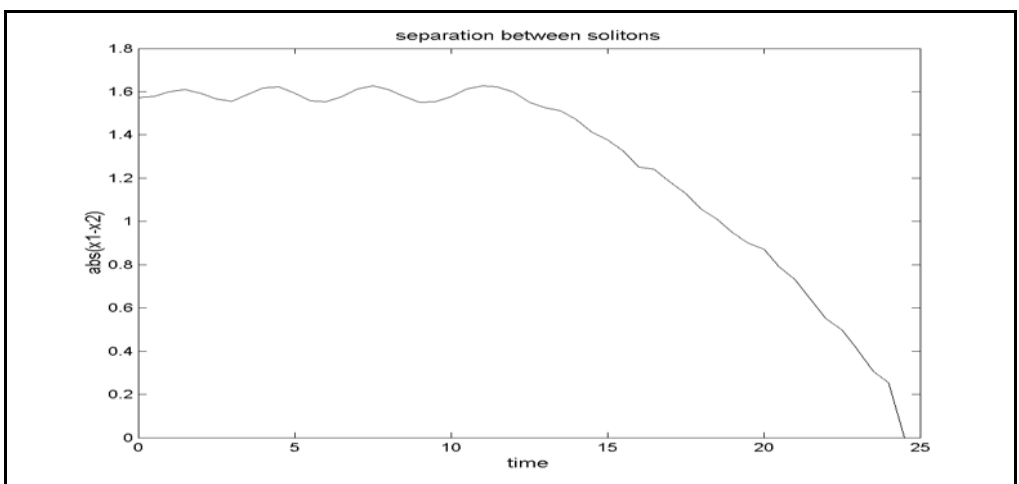

Fig. 11 A plot of the soliton separation as a function of time. From t=0 to about t=12 there is no apparent attraction. Then, the two solitons move with approximately constant relative acceleration until they collide at t=24.5.

**Characteristic curves in the presence of a soliton in 1D**

Let's consider a pure soliton with no noise. If we are in a frame where the soliton is at rest, then the mean fluid velocity is non-zero. There are two points where the fluid velocity is zero, one just upstream and one just downstream of the peak, and any characteristic curve that starts at one of these points is simply a constant value of x independent of time. Now we can consider other starting points, and their characteristic curves asymptote only to the upstream zero-velocity point as shown in figures 12 and 13. This upstream zero-velocity point is thus an attractor for all conserved material quantities that co-move with the fluid. The downstream zero velocity point is unstable. So except for this unstable point, all characteristic curves asymptote to the same limit point. This explains qualitatively at least why two solitons attract. When weak noise is added to the pure soliton, the situation remains essentially the same, except that the characteristic curve has added to it a random walk. The upstream edge of the soliton still acts as an attractor for all characteristic curves.

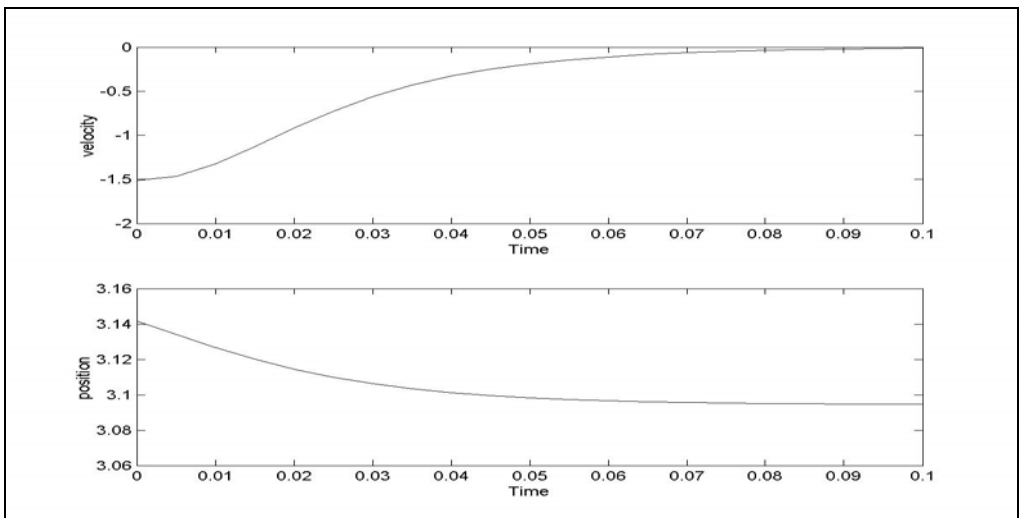

Fig. 12 Soliton centered at x=π and stationary, mean flow is +1, Characteristic starts at center of soliton with negative velocity and moves to the zero velocity point of the soliton on the upstream edge

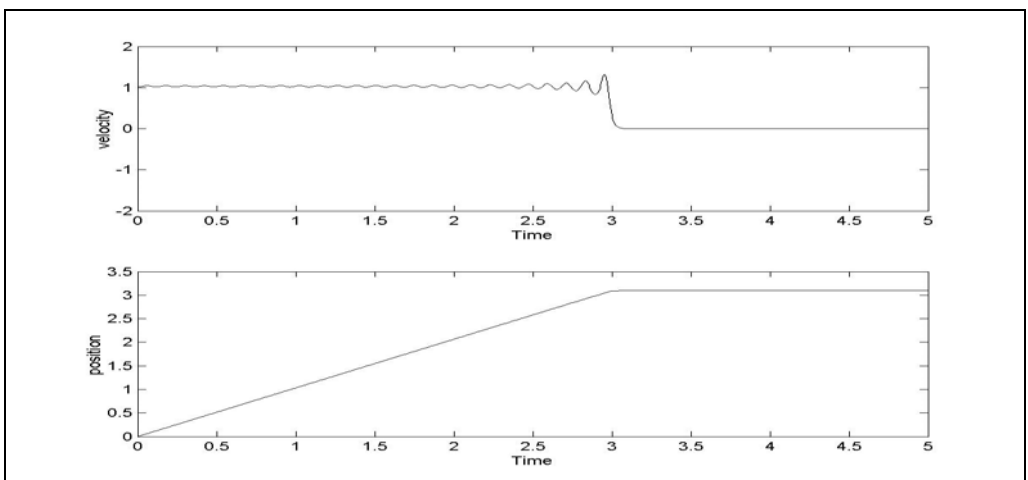

Fig. 13 Soliton centered at x=π and stationary, mean fluid velocity is +1, Characteristic starts at x=0, far away from soliton and moves to the zero velocity point of the soliton on the upstream edge of the soliton

**Motion of solitons under external forces**

We may add an external body force to the fluid equations obtaining

$$\frac{\partial u_\Lambda}{\partial t} + \frac{1}{2}\frac{\partial}{\partial x}P_\Lambda(u_\Lambda^2) = F_\Lambda(x,t) \tag{63}$$

where $F_\Lambda(x,t)$ represents the ratio of force to mass density. Consider the acceleration of a characteristic curve. It would satisfy

$$\frac{d}{dt}x(t) = u(x(t),t) \text{ and } \frac{d^2}{dt^2}x(t) = F_\Lambda(x(t),t) + \frac{1}{2}\frac{\partial}{\partial x}(1-P_\Lambda)u_\Lambda^2 \tag{64}$$

In other words, a characteristic curve would satisfy Newton's equation for an external force, but there would be an addition high spatial frequency force correction, stochastic in nature, due to the truncation or momentum cutoff. Under a weak low spatial frequency force, a soliton would be expected to follow the Newtonian trajectory approximately as well, since it would not change its shape very much and the characteristic curves near the peak would be captives of the soliton. For strong external forces, the soliton would be distorted, and the subsequent motion would be complicated. Simulations confirm these qualitative behaviors.

**Invariant subspaces**

It was noted in [15] that there are invariant subspaces for this system. Observing (42) notice that

> if, at time t, $\hat{u}_k = 0$ unless $k = mk_0$ for $m$ and $k_0$ integers
> with $k_0$ fixed at a single value, then this property holds for  (65)
> all time

Thus the subspace defined by this condition is invariant. Solutions are periodic in x with period $2\pi/k_0$. Solitons can exist in this subspace too. The static soliton condition **(46)** for the invariant subspace becomes

$$\frac{\partial \hat{u}_k}{\partial t} = 0 = \frac{-ik}{2}\sum_{\substack{|k'|,|k-k'|\leq \Lambda \\ |k|\leq \Lambda}} \hat{u}_{k-k'}\hat{u}_{k'} \tag{66}$$

where $k$ and $k'$ are restricted to multiple values of $k_0$. This leads to the same equation for the soliton but with the proviso that $k$ and $k'$ are multiples of $k_0$. The resulting soliton solution is the same as would be found for $k_0 = 1$, but for a reduced value of $\Lambda$ which is defined by the formula

$$\Lambda(k_0) = \Lambda / k_0 \text{ rounded down to the next smallest integer} \qquad (67)$$

These actually describe multiple equally spaced solitons in x with multiplicity $k_0$. Figures 3 appears to be an example of a multiple soliton of this type, but figure 4 does not. They are both probably not stable because a small a small error term in the soliton's shape which is not in the invariant subspace will tend to grow with time. A special case of this invariant subspace is defined by the condition $k_0 > \Lambda/2$, and for this subspace the equations for $\hat{u}_k$ are linear, given by

$$\frac{\partial \hat{u}_k}{\partial t} = \frac{-ik}{2} \sum_{\substack{|k'|,|k-k'|<=\Lambda \\ |k|<=\Lambda}} \hat{u}_{k-k'}\hat{u}_{k'} = -ik\hat{u}_k\hat{u}_0 \text{ if } \hat{u}_k = 0 \text{ for all } k <= \Lambda/2 \qquad (68)$$

These linear solutions correspond to traveling waves with angular frequency $\omega = k\hat{u}_0$ and phase velocity $\omega/k = \hat{u}_0$. Note that linear superposition does not hold for these traveling waves.

**Completeness of the soliton functions**

Let $u_{s,\Lambda}(x)$ denote a stationary soliton solution for cutoff $\Lambda$ and $\hat{u}_0 = 1$ whose peak is at x=0. Then we can expand a general velocity field in the following way

$$u_\Lambda(x_j,t) = \sum_{j'=0}^{2\Lambda} a_{j'}(t) u_{S,\Lambda}(x_j - x_{j'}), \quad x_j = 2\pi j/(2\Lambda+1), \, j=0 \text{ to } 2\Lambda \qquad (69)$$

This coefficients of $a_{j'}(t)$ make up a circulant matrix. Numerical analysis shows that these matrices are non-singular, and therefore the $a_{j'}(t)$ can be solved for in terms of the $u_\Lambda(x_j,t)$. In fact, since the eigenvalues of a circulant matrix comprise the discrete Fourier transform (DFT) of the first row of the matrix [38], it follows that the matrix C below is non-singular since the DFT of $u_{S,\Lambda}$ have only nonzero elements for the cutoff values $\Lambda$ which we have examined (50, 100, 200, 1000, and 5000).

$$C_{jk} = u_{S,\Lambda}(x_j - x_k), \, j,k = 0 \text{ to } 2\Lambda \qquad (70)$$

Thus at any instant of time, it is possible to imagine that the fluid motion is a linear superposition of soliton states at different locations and with different weights without loss of generality. This is vaguely reminiscent of quantum mechanics again, where the eigenstates of the position operator for a particle form a complete set of states.

**The Three dimensional Burgers Equation**

In three dimensions the inviscid Burgers equation becomes

$$\frac{\partial \mathbf{u}}{\partial t} + (\mathbf{u} \cdot \nabla)\mathbf{u} = 0 \tag{71}$$

We assume that $\mathbf{u}$ is periodic in $2\pi$ displacements in all three Cartesian directions. Most studies of this equation assume an incompressibility condition which leads to a conserved energy density proportional to $\mathbf{u}^2$. If the fluid is allowed to be compressible then conserved quantities can be formed in the following way. If $f_G$ is any solution to the conservation equation

$$\frac{\partial f_G}{\partial t} + \nabla \cdot f_G \mathbf{u} = 0 \tag{72}$$

then it follows that the following are invariants

$$\int f_G d^3x, \int f_G \mathbf{u} d^3x, \int f_G \mathbf{u}^2 d^3x \tag{73}$$

provided the integrals exist, and are sufficiently regular. In fact, a general product of the following form is also invariant as follows by induction

$$\int f_G \prod_{j=1}^{N} u_{k(j)} d^3x, \; k(j) \text{ any integer function of j and for all N} \tag{74}$$

In addition to these invariants, the Kelvin circulation theorem applies at least formally, and the following integrals are invariant for any closed curve C, and where C is embedded in and flows with the fluid

$$\Gamma_C = \oint_C \mathbf{u} \cdot d\mathbf{s} \tag{75}$$

Finally, there are topological invariants related to helicity for this system [39]. Defining the helicity by

$$\mathscr{H}(t) = \int_V \mathbf{u} \cdot \boldsymbol{\omega} dV, \quad \boldsymbol{\omega} = \nabla \times \mathbf{u} \tag{76}$$

where the integration volume is bounded by a closed surface S(t) which moves with the fluid, and $\boldsymbol{\omega} \cdot \hat{\mathbf{n}} = 0$ where $\hat{\mathbf{n}}$ is the surface normal. Then for each such volume V, the helicity is an invariant, again at least formally. The topology of the surface cannot change, and thus this result defines a potentially large class of topological invariants as well as a quite literal mechanism for long-range entanglement, another feature of quantum mechanics.

Making rigorous statements about conservation laws in the presence of shocks is mathematically difficult and beyond the scope of this paper, and so we emphasize the formal nature of these invariants for the inviscid Burgers equation.

We can generalize the Fourier-Galerkin spectral method to three dimensions by defining a 3D projection operator

$$P_{\Lambda,\Lambda,\Lambda}\mathbf{f} = P_\Lambda \mathbf{f} = \mathbf{f}_\Lambda = \sum_{\substack{|k_j|<=\Lambda \\ j=1,2,3}} \hat{\mathbf{f}}_\mathbf{k} e^{i\mathbf{k}\cdot\mathbf{x}} \tag{77}$$

$$\hat{\mathbf{f}}_\mathbf{k} = \frac{1}{(2\pi)^3} \int_0^{2\pi}\int_0^{2\pi}\int_0^{2\pi} \mathbf{f}(x) e^{-i\mathbf{k}\cdot\mathbf{x}} d^3x \tag{78}$$

Usually the incompressible three dimensional ideal fluid is studied which has a conserved energy, even in the spectral truncation form [24]. The truncated equation becomes

$$\frac{\partial \mathbf{u}_\Lambda}{\partial t} + P_\Lambda\left[(\mathbf{u}_\Lambda \cdot \nabla)\mathbf{u}_\Lambda\right] = 0 \tag{79}$$

A good source for spectral methods is [40]. We can construct an infinite number of invariants again by letting $(f_G)_\Lambda$ satisfy

$$P_\Lambda\left[\frac{\partial (f_G)_\Lambda}{\partial t} + \nabla \cdot (f_G)_\Lambda \mathbf{u}_\Lambda\right] = 0 \tag{80}$$

and then if follows that
$$\mathbf{P}_f = \int (f_g)_\Lambda \mathbf{u}_\Lambda d^3x \tag{81}$$

is invariant. But an energy invariant is not so obvious. Moreover, there does not appear to be an analog of the invariant H which was found for the 1D inviscid Burgers equation.

**Soliton solutions of the truncated 3D inviscid Burgers equation**

We can generate 3D solitons from the 1D solitons in the following way. Let's first generate a soliton for $\mathbf{u}$ pointing along a Cartesian axis, say $\hat{x}$. Working in the rest frame of the soliton so $\dot{\mathbf{u}} = 0$ we find factorized solutions of the form

$$\mathbf{u}(\mathbf{x}) = u_S(x) u_\perp(y,z)\hat{\mathbf{x}} \tag{82}$$

where $u_S(x)$ is a one dimensional soliton solution. In addition to these solutions are all those derived from them by the combined action of Galilean transformations, translations in time and space, and scale transformations. The spatial form of the narrowest soliton is

$$\mathbf{u}(\mathbf{x}) = u_{1S}(x) \delta_\Lambda(y) \delta_\Lambda(z)\hat{\mathbf{x}} \tag{83}$$

plus all the solutions generated from this by covariance groups, where $\delta_\Lambda(x)$ is the delta function projected onto the truncated Fourier space which is expressible in terms of a sinc function. Solitons moving in directions not parallel to the Cartesian axes also exist at certain special angles.

In 1D the invariance of H led to stability for solitons in the presence of turbulence in the fluid, but in three dimensions H is no longer invariant, and so the stability of 3D solitons in the presence of turbulence must be studied. In the limit of large cutoff Λ, the size of the soliton goes to zero as 1/Λ, and thus the cross-section for scattering of two solitons would also go to zero in this limit, and so, in 3D the solitons would become non-scattering in this limit because the chance of a random collision would be vanishingly small.

**Azimuthal solitons in cylindrical coordinates**

Consider the 3D inviscid Burgers equation in cylindrical coordinates with an external radial force.

$$\dot{\mathbf{u}} + (\mathbf{u} \cdot \nabla)\mathbf{u} = \mathbf{F} \tag{84}$$

We apply a spectral method only to the azimuthal direction here. With $\mathbf{u} = u_\varphi \hat{\varphi}$, and using $\frac{\partial}{\partial \varphi}\hat{\varphi} = -\hat{r}$, the Burgers equation becomes

$$\dot{u}_\varphi \hat{\varphi} + \frac{1}{2r}\frac{\partial}{\partial \varphi}\left(u_\varphi^2\right)\hat{\varphi} - \frac{u_\varphi^2}{r}\hat{r} = F_r \hat{r} \tag{85}$$

In order to have a solution, we must have the special external force satisfying $F_r = -\frac{u_\varphi^2}{r}$. This is an unusual force, but for this special case, the equation becomes

$$\dot{u}_\varphi + \frac{1}{2r}\frac{\partial}{\partial \varphi}\left(u_\varphi^2\right) = 0 \tag{86}$$

Assume a separable solution of the form $u_\varphi = R(r)Z(z)\phi(\varphi,t)$ leads to the equation

$$R(r)Z(z)\dot{\phi}_\varphi + \frac{(R(r)Z(z))^2}{r}\phi_\varphi \frac{\partial}{\partial \varphi}\phi_\varphi = 0 \tag{87}$$

This has one-dimensional soliton solutions of the type studied above in the $\varphi$ coordinate, plus all solutions generated from these by the covariance symmetry groups of the equation. Solutions of this type can have a single period of oscillation provided that $RZ/r = 0$ or 1 for all r and z. These solitons act like a particle with spin. They can be localized in space by choosing $RZ=0$ outside some range, and they have an internal pulse which is the period of oscillation of the soliton. Also, they have a time-averaged rest frame in which their energy is nonzero even if the mean velocity of the fluid is zero in that frame. Note that if *Z(z)* is a constant, then the soliton looks like a tube, and if *R(r)* is limited to small r (ie. *R(r)*=0 for r > r$_{max}$) then this soliton looks like a string.

Besides RZ/r = 1, any other constant would do as well, and so the angular velocity around the orbit of the soliton is arbitrary.

**Solitons in higher dimensional geometries**

Here we consider a fluid which flows in a geometry with extra compactified dimensions as in Kaluza-Klein theory [41,42] or string theory [43,44]. For simplicity let us first consider one extra space dimension. The inviscid Burgers equation is trivial to generalize. All that is required is to use the N dimensional gradient in it. Let's consider for simplicity a single extra dimension. Let the extra compactified dimension be denoted by $x_4$. We require periodicity in this extra dimension so that $\mathbf{u}(\mathbf{x},t,x_4+2\pi r) = \mathbf{u}(\mathbf{x},t,x_4)$, where r is a constant. We look for a soliton solution in the $x_4$ dimension of the form $\mathbf{u} = f(x_1,x_2,x_3)u_s(x_4/r,t/r)\hat{x}_4$ where $u_s$ is a soliton solution. Then the equation becomes

$$f\dot{u}_s(x_4/r,t/r) + f^2 u_s(x_4/r,t)\frac{\partial}{\partial x_4}u_s(x_4/r,t/r) = 0 \tag{88}$$

When spectrally truncated, this equation will be a solution provided that $f = 0$ or 1 at each 3 dimensional point **x.** From these solutions many more can be generated by applying the covariance groups of the inviscid Burgers equation for this 4 space dimensional system. Again, stability in the presence of turbulence will need to be examined for these. Solitons moving in more than one compactified dimension can also be constructed.

**Possible relationship to string theory**

Moffat's topological invariants [39] provide a mechanism for the stability of closed tubes of vorticity. If these tubes are narrow in cross section, then they look like strings, and if the initial conditions are such that the fluid starts off like this, then this string resemblance should persist for some period of time. There is an interesting connection between string theory and magnetohydrodynamics with narrow magnetic flux tubes, and also with vortex flux tubes in superfluid turbulence theory [45-47]. The observed stochastic vortex tangle in superfluid turbulence He II suggests a similar phenomenon should occur in the inviscid Burgers system. In fact, if the inviscid ideal fluid (in 3D of course) started off with just a single filamentary vortex, ie. a single stringlike tube of vorticity, then as time progressed this single flux tube could become extraordinarily entangled, filling the whole fluid volume with a tangle of vorticity, but every single piece of vortex tube would still have the same flux as the starting vortex tube. This is due to the Kelvin circulation theorem. Thus the Kelvin circulation around any loop would be quantized in essentially the same manner as in a superfluid. This is a potential candidate for a mechanism underlying quantum entanglement in a theory such as is being proposed here to try and derive quantum mechanics from a fluid model.

**The fluid as a Markov process**

The truncated inviscid Burgers equation is first order in time and by Picard's existence theorem it has a unique solution for any initial conditions. Therefore, the state of the entire fluid at one instant in time determines the fluid for all times. Consequently any statistical treatment of the fluid will be a Markov process in both the forward and backward time directions. This is similar to Nelson's stochastic mechanics [48,49] which is based on a dynamical assumption which is time reversible and stochastic. Since the inviscid Burgers equation is time reversible and inherently stochastic it is remarkably similar to Nelson's theory. One of the problems with Nelson's theory was always that the main examples of Brownian motion known in physics were due to highly dissipative systems which are not time reversible. In our present work we have a system which achieves a stochastic behavior and Markov behavior without dissipation. Stochastic mechanics assumes, on the microscopic level, that the trajectories look locally like Wiener processes. This is clearly not true for the fluid, except as some coarse time approximation.

The motion of soliton peaks or of characteristic curves on the other hand can be derived from the fluid, but they are not necessarily Markov processes except in some approximation. In the presence of ergodic noise, the soliton's peak becomes effectively a random variable that depends on the fluid. Thus the soliton's peak trajectory will be describable by a hidden Markov model [50]. Stochastic mechanics is known to be highly non-unique [51,52], and therefore it is conceivable that it could be equivalent to such a hidden Markov model in some approximation.

## Quantization

The single valuedness of the quantum mechanical wave function is not an automatic property of stochastic models or hydrodynamic models [53,54]. We now propose a mechanism to explain this phenomenon in the current context. Here we are not restricting consideration to a Fourier cutoff theory. Suppose that the cosmological proto fluid started out with one infinitesimally thin stringlike vortex tube. This would be an initial condition for the entire universe. The vortex tube might be closed, or open and infinite. The rest of the space outside the vortex tube we assume for the time being had zero vorticity. Then, over time, assuming the fluid were turbulent, this single vortex tube would become immensely tangled and confused, but its flux would be maintained provided there were only finite external forces applied to the fluid. The reason the flux would be maintained is that since the vortex is infinitesimal, it follows that for an infinitesimal loop c

$$\frac{d}{dt}\oint_c \mathbf{u}\cdot \mathbf{dl} = O(\oint_c dl) \text{ as } \oint_c dl \to 0 \tag{89}$$

provided that $|\mathbf{u}|$, $\left|\frac{D\mathbf{u}}{Dt}\right|$, and $|\nabla \mathbf{u}|$ are bounded

In developing a model based on this idea we might allow vortex crossing and recombination effects [55] to occur which would allow closed loops of thin vortex tubes to separate from the main string and go their own way inside the fluid while inheriting the flux of the parent vortex. These topological solitons might persist and interact. Assuming that an arbitrary closed curve C does not exactly intersect any vortex string we have that

$$\oint_C \mathbf{u}\cdot \mathbf{dl} = N\kappa, \text{ for } N \text{ integer and } \kappa \text{ a constant} \tag{90}$$

where $\kappa$ is the universal flux of all the vortex strings and N depends on how many vortex strings are linked by the curve C and on their orientation.

In this situation, for any smooth and simply connected domain $\Omega$ which does not contain any sections of any vortex strings we can write

$$\mathbf{u} = \nabla S, \, x \in \Omega \tag{91}$$

This is the weak form of the Helmholtz theorem. By taking overlapping domains we can analytically continue S to the whole space. But S will not be single valued because if it were then the circulation would vanish for all closed curves. The multiple values of S can be $\{S + l\kappa, \, l \text{ integer}\}$. Because S is not single valued, it is not convenient mathematically to describe the fluid in terms of it. However, note that $e^{\pm iS2\pi/\kappa}$ is single valued. Therefore there is a real advantage to transforming the fluid equations into complex ones to deal with single valued functions and still have the simplicity of a potential fluid. Let us now consider a conserved quantity moving in the fluid. It might be mass, or charge, or the probability density of a diffusion. The conservation equation reads

$$\frac{\partial \rho}{\partial t} + \nabla \cdot \rho \nabla S = 0 \tag{92}$$

and the equation of motion for the inviscid Burgers fluid becomes

$$\frac{\partial}{\partial t}\nabla S + (\nabla S \cdot \nabla)\nabla S = \frac{\partial}{\partial t}\nabla S + \frac{1}{2}\nabla (\nabla S)^2 = 0 \tag{93}$$

These two equations are equivalent to the following Schrödinger type equation by a Madelung transformation

$$\left[-\frac{1}{2}\left(\frac{\kappa}{2\pi}\right)^2 \Delta + V_\kappa\right] e^{R+iS2\pi/\kappa} = i\left(\frac{\kappa}{2\pi}\right)\frac{\partial}{\partial t} e^{R+iS2\pi/\kappa} \text{ where } \rho = e^{2R} \quad (94)$$

where

$$V_\kappa = \frac{1}{2}\left(\frac{\kappa}{2\pi}\right)^2 \frac{\Delta e^R}{e^R} \quad (95)$$

defining a "wave function" by

$$\Psi = e^{R+iS2\pi/\kappa} \quad (96)$$

we have a Schrödinger-like equation except that the potential given by (95) is nonlinear. The single-valuedness of this wave function is thus a consequence of the starting conditions for the fluid which had of a single infinitely thin vortex tube.

We may generalize this result to include not only the singular vortex string, but also a smooth vorticity in addition. In this case, we may use the weak Helmholtz theorem again to write at points not on a vortex string as

$$\mathbf{u} = \mathbf{A} + \nabla S \quad (97)$$

where $\mathbf{A}$ is a smooth and single-valued vector function which has nonvanishing curl somewhere. It is defined only up to a gauge transformation.

$$\mathbf{A} \to \mathbf{A} + \nabla \xi(\mathbf{x},t) \text{ and } S \to S - \xi(\mathbf{x},t) \text{ leaves } (A+\nabla S) \text{ invariant} \quad (98)$$

Choose a gauge so that $\nabla \cdot \mathbf{A} = 0$, analogous to the Coulomb gauge. The fluid equation is

$$\dot{\mathbf{A}} + \nabla \dot{S} + ((\mathbf{A}+\nabla S)\cdot \nabla)(\mathbf{A}+\nabla S) = 0 \quad (99)$$

we have the identity

$$(\mathbf{u}\cdot\nabla)\mathbf{u} = \frac{1}{2}\nabla \mathbf{u}^2 - \mathbf{u}\times(\nabla\times\mathbf{u})$$
$$\Rightarrow \dot{\mathbf{A}} + \nabla\dot{S} + \frac{1}{2}\nabla(\mathbf{A}+\nabla S)^2 - (\mathbf{A}+\nabla S)\times(\nabla\times\mathbf{A}) = 0 \quad (100)$$

Now we do a Helmholtz decomposition on the rightmost term into a solenoidal and rotational part

$$(\mathbf{A}+\nabla S)\times(\nabla\times\mathbf{A}) = -\nabla\Phi + \nabla\times\Omega \quad (101)$$

Then we obtain the following two equations

$$\dot{\mathbf{A}} - \nabla \Omega = 0 \tag{102}$$

and a Hamilton-Jacobi equation

$$\dot{S} + \frac{1}{2}(\mathbf{A} + \nabla S)^2 + \Phi = 0 \tag{103}$$

The conservation equation is

$$\frac{\partial \rho}{\partial t} + \nabla \cdot \rho (\nabla S + \mathbf{A}) = 0 \tag{104}$$

Then the following Schrodinger-like equation is equivalent to these equations as is shown again by setting real and imaginary parts to zero in a Madelung transform:

$$\left[ -\frac{1}{2}\left(\frac{\kappa}{2\pi}\right)^2 \left(\nabla + i\frac{2\pi}{\kappa}\mathbf{A}\right)^2 + \Phi + V_\kappa \right] e^{R+iS\frac{2\pi}{\kappa}} = i\left(\frac{\kappa}{2\pi}\right)\frac{\partial}{\partial t} e^{R+iS\frac{2\pi}{\kappa}} \text{ where } \rho = e^{2R} \tag{105}$$

This bears a formal resemblance to the minimal coupling magnetic force Schrödinger equation, except for the nonlinear potentials $\Phi$ and $V_k$. It has the same U(1) abelian gauge invariance too, where **A** plays the role of a pseudo vector potential. But here there is no external electromagnetic field. The equivalence of electromagnetism and turbulent ideal fluids has been studied before for incompressible fluids [56,57]. Whether such a formal equivalence can be made for compressible systems as studied here is not clear. For quantum mechanics viewed as fluid equations, the velocity field depends on the mass of the particle. In our equation, which might apply to the early universe, there are no particles, only a fluid. Particles come later, at least the particles that we observe in a laboratory. Explaining how the masses of the particles get into the Schrödinger equation will not be addressed here, and might be a difficult problem for this picture in the future. For now we are content to have derived a Schrödinger-like equation for the fluid in the early universe, and to have presented numerical models which exhibit time reversible chaos.

The extra terms $\Phi$ and $V_\kappa$ in (105) result in a nonlinear Schrödinger equation. Note that $V_\kappa$ depends only on the density whereas $\Phi$ depends only on the fluid velocity fields. They need to be cancelled somehow if we are to have linear superposition and Schrödinger's equation. Let's consider irrotational flow (except for the vortex strings) so that **A**=0 and consequently $\Phi = 0$. If the fluid started off irrotational, it would remain so unless acted upon by a

rotational force. One way to cancel $V_\kappa$ would be to add a body force acting on the fluid of the form

$$V_{QM} = -V_k = -\frac{1}{2}\left(\frac{\kappa}{2\pi}\right)^2 \frac{\Delta\sqrt{\rho}}{\sqrt{\rho}} \tag{106}$$

Historically this force has been called the quantum mechanical potential. It is not sufficient to simply posit the existence of such a force in an ad hoc manner. It must come from somewhere, and ultimately it must be derived rigorously from some physical principle. It is the author's opinion that this extra term may be related to radiation reaction effects. The only derivation he is aware of for this strange type of force was given in [58], and it involved a radiative reactive force in thermal equilibrium with a statistical medium like our present fluid in spectral truncation and exhibiting ergodic behavior. Here is another reason that suggests there is a connection with radiation. Table 3 shows the Larmor formula for bremsstrahlung calculated in different ways for charged quantum and classical particles taken from [59]

| Classical radiation result for a point particle | $E_{rad} = \frac{2}{3}\frac{q^2}{c^3}\int_0^T \mathbf{a}^2(t')dt'$ |
|---|---|
| Hydrodynamic model using classical electromagnetism | $E_{rad} = \frac{2}{3}\frac{q^2}{c^3}\int_0^T \left|\langle\Psi|\mathbf{a}(t')|\Psi\rangle\right|^2 dt'$ |
| Conventional Quantum Radiation Result | $E_{rad} = \frac{2q^2}{3c^3}\int_0^T \langle\Psi|\mathbf{a}^2(t')|\Psi\rangle dt'$ |

Table 3 Larmor formulae for bremsstrahlung

Note that the hydrodynamic classical result has the square of the expectation of the acceleration vector, but the quantum result has the expectation of the acceleration squared. Consider the acceleration which would be caused by the potential term $V_\kappa$ in these expressions if $\rho$ described a charge density. If we use classical electromagnetism, the expression for radiation is zero

$$E_{rad \atop classical} \propto \int_0^T \left[\int \rho(x)\nabla V_\kappa d^3x\right]^2 dt' \propto \int_0^T \left[\int \rho(x)\nabla\left(\frac{\Delta\sqrt{\rho}}{\sqrt{\rho}}\right)d^3x\right]^2 dt' = 0 \tag{107}$$

which follows simply by integration by parts. This is perfectly consistent with the fact that the Burgers fluid has characteristic curves which are straight lines moving at constant velocity if we don't perform spectral truncation and if there are no shocks. This doesn't show any necessity for a quantum mechanical potential. But the quantum mechanical radiation is not zero if we treat our wave function as if it were the quantum mechanical one.

$$E_{rad \atop QM} \propto \int_0^T \left[ \int \rho(x) |\nabla V_\kappa|^2 \, d^3x \right] dt'$$
$$= \int_0^T \left[ \int \rho(x) \left| \nabla \left( \frac{\Delta\sqrt{\rho}}{\sqrt{\rho}} \right) \right|^2 d^3x \right] dt' \geq 0 \qquad (108)$$

The only way for the quantum mechanical result to be zero is if $|\mathbf{a}| = |\nabla(V_\kappa + V_{QM})| = 0$ at least everywhere that $\rho \neq 0$. This requires that $V_{QM} = -V_\kappa$ which is exactly what is required to get Schrödinger's equation. The characteristic curves would no longer be straight lines in this case, but would be Bohmian trajectories which in general have curvature, even for a free particle.
.

So, if we could justify using the quantum mechanical bremsstrahlung formula, then we could make a reasonable argument in support of the quantum mechanical potential. But why should the radiation be different from the classical result for our fluid? One reason is possibly that shocks must be considered. Shocks are well known to be an important feature of the Burgers equation [60,61]. Perhaps the correct inclusion of bremsstrahlung classically involves fluid shocks in this case. Suppose that at some set of points in spacetime there are shocks, and we use the classical hydrodynamic model. The acceleration at a shock point will be infinite, and so

$$E_{rad \atop classical} \propto \int_0^T \left[ \int \rho(x,t) \sum_m \mathbf{a}_m(x,t) d^3x \right]^2 dt' \qquad (109)$$

will in general be nonzero where the $\mathbf{a}_m$ are distributions or singular functions about the points $x_m$ where shocks occur. Since shocks occur seemingly randomly, there is no reasonable expectation of cancellation in this expression, and so the resulting classical radiation will be non-zero. This is a relaxation mechanism. The fluid will tend to respond by minimizing this radiation. But notice what happens if we include a quantum mechanical potential such that $V_{QM} = -V_\kappa$. The equation for the fluid including $V_{QM}$ becomes

$$\left[ -\frac{1}{2} \left( \frac{\kappa}{2\pi} \right)^2 \Delta + (V_\kappa + V_{QM}) \right] e^{R+iS\frac{2\pi}{\kappa}} = i \left( \frac{\kappa}{2\pi} \right) \frac{\partial}{\partial t} e^{R+iS\frac{2\pi}{\kappa}} \qquad (110)$$

and then if $V_{QM} = -V_\kappa$ we get

$$\left[ -\frac{1}{2} \left( \frac{\kappa}{2\pi} \right)^2 \Delta \right] e^{R+iS\frac{2\pi}{\kappa}} = i \left( \frac{\kappa}{2\pi} \right) \frac{\partial}{\partial t} e^{R+iS\frac{2\pi}{\kappa}} \qquad (111)$$

Now we have a remarkable result. The nonlinear fluid equations have been linearized, but more importantly **there are no more shocks!** And also there is no more radiation for the current for the free Schrödinger equation does not radiate classically [62]. The inclusion of the quantum mechanical potential $V_{QM}$ has acted as a shock absorber and a radiation supressor. So we see how electromagnetic radiation could possibly suppress shocks, and a very effective way that this could be achieved is if the effect of the radiative reaction is to cause a quantum mechanical potential to appear which produces a force on the fluid which eliminates shocks. Granted this is far from a proof, but at least it is something to ponder, and it gives a physical principle which could be the source of the quantum mechanical potential.

The shock suppression argument leads to another intriguing possibility. We have assumed in this section that the fluid satisfied vorticity flux quantization due to the existence of a single vortex string at the beginning of the universe. And this led rather directly to a Schrödinger-like equation. And then we found that if the quantum mechanical potential were present, it would suppress shocks and thereby suppress electromagnetic radiation. So in other words, the suppression of shocks requires both flux quantization and the quantum mechanical potential acting on the fluid. Is it possible that flux quantization could be an emergent phenomenon too, caused by the relaxation of the fluid to radiation produced by shocks? If this were the case then it wouldn't be necessary to start the fluid off with a single vortex string. The interaction with radiation could cause the fluid to relax into a state in which vorticity was quantized since otherwise shocks in the fluid could not be prevented. Any form of radiation: acoustic, electromagnetic, gravitational, or non-abelian gauge radiation could presumably serve as the relaxation mechanism to dampen shocks. In the author's opinion, it is quite possible that some of the mathematical methods developed by Adler and coworkers [63] may be relevant to this relaxation problem, but the precise relationship has not yet been established. The main problem is to interpret the non-commuting matrices in trace dynamics in terms of classical stochastic fluids.

Fluid shocks may play still another role in this theory. Because of vorticity quantization, the inviscid Burgers fluid cannot change continuously from one state to another, say on emission of radiation due to some external force applied to the fluid, without transiently violating vorticity quantization. Thus the fluid shocks might be a mechanical explanation for spontaneous wave function collapse [64,65]. In fact there is just such a spontaneous wave function collapse model which is based on Burgers equation [22]. So the historical and popular concept of a quantum jump actually might have a mechanical explanation in this picture. As photon emission or absorption could be accompanied by a fluid shock too, this might help to resolve some profoundly paradoxical aspects of photons.

We have discussed electromagnetic radiation as a possible explanation of the appearance of the quantum mechanical potential, but other types of radiation might have a similar effect, such as acoustic, Yang-Mills, or gravitational radiation.

## Bell's Theorem

Bell's theorem [66] does not apply to a deterministic theory since all physical variables in such a theory are determined by the initial conditions. The "free will axiom" is required for any conclusions to be made regarding nonlocality and quantum mechanics [67]. The fluid model we are considering is deterministic, although because of turbulence it is essentially incalculable for any significant time interval. Critics of hidden variables often comment that a deterministic model of quantum mechanics is physically unacceptable because it is obvious that we have free will, even though they also usually admit, albeit reluctantly, that logically speaking deterministic models do avoid the application of Bell's theorem and the conclusion that a realistic model of quantum mechanics is nonlocal. But is it so obvious that we actually have free will? No experiment has ever been proposed that would test the "free will axiom". The property of free will seems to be unmeasurable. Moreover, analysis of human decision making in terms of collective neural interactions suggests that a conscious decision is a very complex collective biological activity. What is actually making a conscious decision? Could not the idea that that we act freely and of our own will be an illusion? Letting our bias towards free will block research in hidden variable models of quantum mechanics seems unwise, especially now that the RHIC experiments have revealed our likely turbulent and classical fluid beginnings.

## Quantum computers

Another argument one hears against a deterministic model of quantum mechanics is that it implies that one could then simulate a quantum computer using a digital computer programmed to simulate the deterministic system, and therefore achieve the same benefits as quantum computing which is thought to be impossible. The fluid turbulence model is highly chaotic, and as a consequence it has many positive Lyapunov exponents. Simulating such a system with accuracy is extremely difficult, and would ultimately require a numerical precision that grows linearly with the time duration of the desired simulation, and in addition would have an extraordinarily large number of degrees of freedom. It seems very unlikely then that such a simulation could provide a means to compete with the performance of a quantum computer.

## Conclusion

A model for an ideal pressureless fluid which has properties similar to quantum mechanics, has been presented. This model fits into a relativistic cosmological framework, and is motivated by the quark-gluon plasma experiments at RHIC. We have shown how the Reynolds number can be expected to be very large for this classical fluid in the early universe. The model has spontaneous chaos together with solitons which act quite literally as localized waves which have particle properties and also can, in combination with larger scale flows, guide point test particles much like de Broglie's pilot wave theory. It also has self-generating chaotic and ergodic turbulence which makes the motion of both the pilot soliton and a particle embedded in it stochastic. The basic equations are time reversible and there is no dissipation of energy in the system studied and so at least in the truncated case in 1D the turbulence persists indefinitely.

A class of localized solitons have been presented , and extended to higher dimensions. These solitons have a number of interesting features. They trap and guide fluid particles near their peak. They maintain their identity in the presence of ergodic noise, and their motion has a random component added to the constant velocity in this case. They seem to attract one another. They damage each other when colliding, and their energy is proportional to their velocity squared in the mean rest frame of the fluid, like a Newtonian particle's kinetic energy.

If the fluid universe model has at it's start one stringlike vortex tube, then the motivation for a complex wave equation is apparent, since this makes the scalar potential description of the fluid velocity field single-valued by transforming the equations into a Schrödinger type one whose wave function is single-valued. The model does not automatically explain linear superposition though for this wave function. Arguments are presented that suppression of fluid shocks is the key to understanding the appearance of a quantum mechanical potential and the subsequent linearization of the theory. Such suppression might be due to radiative reaction forces on the fluid if it contained charge, or to acoustic or Yang-Mills gauge theory radiation, or possibly even to gravitational radiation. A rigorous derivation of the quantum mechanical potential is not presented, however, and remains a major item for future work.

Based on these limited successes, it is proposed that this type of turbulence be studied further with a goal of finding a variation of the approaches taken here which leads to a still closer agreement with quantum mechanics.

Due to the topological invariants of inviscid fluids there is a potential connection between string theory and turbulence provided that the initial vorticity is confined to stringlike filaments, or that the fluid relaxes into such a state due to some form of radiation. As this filament gets tangled up over time due to turbulence, the Kelvin circulation around any loop remains quantized, acting in this way like a superfluid. Thus this theory may complement string theory and perhaps open some new avenues for analysis.

The question arises as to how one should include gauge fields into the current framework. The usual way is simply to add magnetohydrodynamic terms to the stress-energy tensor, or to add their analog in non-abelian gauge theories. But it's also worth considering that some gauge fields might themselves be emergent properties of a cosmic turbulent fluid.

The possibility that quantum mechanics might be understood at a deeper level from data to be gathered in future heavy ion experiments at CERN and possibly also at an enhanced RHIC may generate added enthusiasm for funding these facilities, and offers the possibility that future predictions of this theory may lead to testable experimental verification.

## Acknowledgements

The author thanks Paul Romatschke for helpful correspondence and Peter Schattner, Andrew Davidson, Michael Jordan, Andrei Khrennikov, Theo Nieuwenhuizen, John Barker, Luis de la Pena, and Ana Maria Cetto for stimulating discussions.

# References

1. E. V Shuryak, "Why does the quark-gluon plasma at RHIC behave as a nearly ideal fluid?," Progress in Particle and Nuclear Physics 53, 273 (2004).
2. H. Song and U. W. Heinz, "Suppression of elliptic flow in a minimally viscous quark-gluon plasma," http://arxiv.org/abs/0709.0742
3. D. Teaney, "The Effect of Shear Viscosity on Spectra, Elliptic Flow, and HBT Radii," http://arxiv.org/abs/nucl-th/0301099
4. J. Noronha, G. Torrieri, and M. Gyulassy, "Near Zone Navier-Stokes Analysis of Heavy Quark Jet Quenching in an $N$ =4 SYM Plasma," http://arxiv.org/abs/0712.1053
5. P. Romatschke and U. Romatschke, "Viscosity Information from Relativistic Nuclear Collisions: How Perfect is the Fluid Observed at RHIC?," http://arxiv.org/abs/0706.1522
6. F. Cooper, G. Frye, and E. Schonberg, "Landau's hydrodynamic model of particle production and electron-positron annihilation into hadrons," Physical Review D 11, 192 (1975).
7. P. Steinberg, "Landau Hydrodynamics and RHIC Phenomena," nucl-ex/0405022 (2004).
8. P. A. Steinberg, "Bulk Dynamics in Heavy Ion Collisions," Nuclear Physics A 752, 423 (2005).
9. E. Madelung," Z. Phys. 40, 322 (1926).
10. L. Susskind, "The Anthropic Landscape of String Theory," http://arxiv.org/abs/hep-th/0302219
11. A. Leggett, "Nonlocal Hidden-Variable Theories and Quantum Mechanics: An Incompatibility Theorem," Foundations of Physics 33 (10), 1469 (2003).
12. S. Groblacher, T. Paterek, R. Kaltenbaek et al., "An experimental test of non-local realism," Nature 446 (7138), 871 (2007).
13. A. Khrennikov, "On the problem of hidden variables for quantum field theory," Nuovo Cimento B 121B, 505 (2006).
14. G. 't Hooft, "Determinism Beneath Quantum Mechanics," in Proceedings of Quo Vadis Quantum Mechanics, Temple University, Philadelphia Pennsylvania, (2002).
15. A. J. Majda and I. Timofeyev, "Remarkable statistical behavior for truncated Burgers-Hopf dynamics," Proc. Natl. Acad. Sci. USA 97 (23), 12413 (2000).
16. R. V. Abramov, G. Kovacic, and A. J. Majda, "Hamiltonian Structure and Statistically Relevant Conserved Quantities for the Truncated Burgers-Hopf Equation," Communications in Pure and Applied Mathematics LVI, 0001 (2003).
17. R. V. Abramov, *Thesis: Statistically relevant and irrelevant conserved quantities for the equilibrium statistical description of the truncated Burgers-Hopf equation and the equations for barotropic flow*, Rensselaer Polytechnic Institute, 2002.
18. A. M. Polyakov, "Turbulence without pressure," Physical Review E 52 (6) (1995).

19. A. M. Polyakov, "The theory of turbulence in two dimensions," Nuclear Physics B 396, 367 (1993).
20. V. P. Dmitriyev, "Mechanical analogy for the wave-particle: helix on a vortex filament," Journal of Applied Mathematics 2 (5), 241 (2002).
21. E. T. Whittaker, in *A History of the Theories of Aether & Electricity: The Classical Theories/the Modern Theories 1900-1926: Two Volumes Bound as One* (Dover, 1990).
22. L. F. Santos and C. O. Escobar, "Burgers turbulence and the continuous spontaneous localization model," Europhysics Letters 54 (1), 21 (2001).
23. P. Romatschke, "Fluid turbulence and eddy viscosity in relativistic heavy-ion collisions," http://arxiv.org/abs/0710.0016
24. C. Cichowlas, Bonaiti. P., F. Debbasch et al., "Effective Dissipation and Turbulence in Spectrally Truncated Euler Flows," Physical Review Letters 95, 264502 (2005).
25. P. F. Kolb and U. Heinz, edited by R. C. Hwa and X.-N. Wang (World Scientific Publishing, Singapore, 2004).
26. R. Baier and P. Romatschke, "Causal viscous hydrodynamics for central heavy-ion collisions," European Physics Journal C 51, 677 (2007).
27. S. Weinberg, *Gravitation and Cosmology: Principles and Applications of the General Theory of Relativity*. (John Wiley and sons, New York, 1972).
28. G. Ellis and H. van Elst, in *Theoretical and Observational Cosmology (also arXiv:gr-qc/9812046v4)*, edited by Marc Lachi (Kluwer, Dordrecht, 1999).
29. J. Kratochvil, A. Linde, E. V. Linder et al., "Testing the cosmological constant as a candidate for dark energy," Juornal of Cosmology and Astroparticle Physics 007, 001 (2004).
30. T. Uchida, "Theory of force-free electromagnetic fields. I. General theory," Physical Review E 56, 2181 (1997).
31. J. M. Burgers, "Mathematical examples illustrating relations occurring in the theory of turbulent fluid motion," Verhand. Kon. Neder. Akad. Wetenschappen, Afd. Natuurkunde, 17, 1 (1939).
32. J. Bec and K. Khanin, "Burgers Turbulence," http://arxiv.org/abs/0704.1611
33. A. J. Majda and I. Timofeyev, "Statistical Mechanics for Truncations of the Burgers-Hopf Equation: A Model for Intrinsic Stochastic Behavior with Scaling," Milan Journal of Mathematics 70, 39 (2002).
34. U. Frisch, *Turbulence: The Legacy of A. N. Kolmogorov*. (Cambridge UniversityPress, 1995).
35. C. O. R. Sarrico, "New solutions for the one-dimensional nonconservative inviscid Burgers equation," J. Math. Anal. Appl. 317, 496 (2008).
36. V. P. Maslov and V. A. Tsupin, "Necessary conditions for the existence of infinitely narrow solitons in gas dynamics," Soviet Phys. Dokl. 24, 354 (1979).

37. V. P. Maslov and O. A. Omel'yanov, "Asymptotic soliton-form solutions of equations with small dispersion," Russian Math. Surveys 36, 73 (1981).
38. R. M. Gray, *Toeplitz and Circulant Matrices: A Review*. (Now, 2005).
39. H. K. Moffatt and A. Tsinober, "Helicity in laminar and turbulent flow," Annual Review of Fluid Mechanics 24, 281 (1992).
40. C. Canuto, M. Y. Hussaini, A. Quarteroni et al., *Spectral Methods, Fundamentals in Single Domains*. (Springer, 2006).
41. T. Kaluza, "On the problem of unity in physics," Sitzungsber. Preuss. Akad. Wiss. Berlin (Math. Phys.), 966 (1921).
42. O. Klein, "Quantum theory and five dimensional relativity," Z. Phys. 37, 895 (1926).
43. J. Polchinski, *String Theory Volumes 1 and 2*. (Cambridge University Press, Cambridge, 1998).
44. M. Kaku, *Hyperspace A Scientific Odyssey through Parallel Universes, Time Warps, and the Tenth Dimension*. (Oxford University Press, Oxford, 1995).
45. P. Olesen, "Dual Strings and Magnetohydrodynamics," Phys. Lett. B366, 117 (1996).
46. S. K. Nemirovskii and W. Fiszdon, "Chaotic quantized vortices and hydrodynamic processes in superfluid helium," Reviews of Modern Physics 67, 37 (1995).
47. V. S. Semenov and L. V. Bernikov," Sov. Phys. JETP 71, 911 (1990).
48. E. Nelson, *Dynamical theories of Brownian motion*. (Princeton University Press, Princeton, 1967).
49. E. Nelson, *Quantum Fluctuations*. (Princeton University Press, Princeton, 1985).
50. L. R. Rabiner, "A Tutorial on Hidden Markov Models and Selected Applications in Speech Recognition," Proceedings of the IEEE 77, 257 (1989).
51. M. Davidson, "A dynamical theory of Markovian diffusion," Physica A 96, 465 (1979).
52. M. Davidson, "A generalization of the Fenyes-Nelson stochastic model of quantum mechanics," Letters in Mathematical Physics 3, 271 (1979).
53. T. C. Wallstrom, "Inequivalence between the Schrödinger equation and the Madelung hydrodynamic equation," Phys. Rev. A 49 (3), 1613 (1993).
54. T. Takabayasi, "On the Formulation of Quantum Mechanics associated with Classical Pictures," Progress of Theoretical Physics 8 (2), 143 (1952).
55. M. Bou-Diab, M. J. W. Dodgson, and G. Blatter, "Vortex Collisions: Crossing or Recombination?," Phys. Rev. Lett. 86, 5132 (2001).
56. H. Marmanis, *Analogy between the Electromagnetic and Hydrodynamic Equations: Applications to Turbulence*, Brown University, 2000.
57. H. Marmanis, "On the analogy between electromagnetism and turbulent hydrodynamics," arXiv:hep-th/9602081v1
58. M Davidson, "A model for the stochastic origins of Schrödinger's equation," Journal of Mathematical Physics 20 (9), 1865 (1979).

59. M. Davidson, "Predictions of the hydrodynamic interpretation of quantum mechanics compared with quantum electrodynamics for low energy bremsstrahlung.," Annales de la Fondation Louis de Broglie, (arxiv.org:quant-ph/0302045 corrects some typesetting errors). 29 (4), 661 (2004).
60. Z. She, E. Aurell, and Uriel Frisch, "The Inviscid Burgers Equation with Initial Data of Brownian Type," Commun. Math. Phys. 148, 623 (1992).
61. Y. G. Sinai, "Statistics of Shocks in Solutions of Inviscid Burgers Equation," Commun. Math. Phys. 148, 601 (1992).
62. M. P. Davidson, "Quantum wave equations and non-radiating electromagnetic sources," Annals of Physics 322 (9), 2195 (2007).
63. S. L. Adler, *Quantum Theory as an Emergent Phenomenon*. (Cambridge University Press, Cambridge, 2004).
64. P. Pearle, "Reduction of statevector by a nonlinear Schrödinger equation," Physical Review D 13, 857 (1976).
65. G. C. Ghirardi, A. Rimini, and T. Weber, "Unified dynamics for microscopic and macroscopic systems," Physical Review D 34, 470 (1986).
66. J. S. Bell, *Speakable and unspeakable in quantum mechanics*. (Cambridge University Press, Cambridge, England, 1987).
67. G. 't Hooft, "The Free-Will Postulate in Quantum Mechanics," http://arxiv.org/abs/quant-ph/0701097v1